\documentclass[prx, 10pt, twocolumn, floatfix, superscriptaddress, aps, longbibliography]{revtex4-1}

\usepackage{times, mathrsfs, amsmath, amsfonts, graphics, graphicx, cancel, color, amsthm, bbm, mathtools, amssymb, physics}

\usepackage[nice]{nicefrac}
\usepackage[english]{babel}
\usepackage[margin=.75in]{geometry}
\usepackage[T1]{fontenc} 
\usepackage{tabularx,ragged2e,booktabs}
\usepackage{capt-of}
\newcolumntype{C}[1]{>{\Centering}m{#1}}

\usepackage{comment}

\usepackage{xfrac,relsize,float}
\usepackage{appendix}
\usepackage{siunitx}
\usepackage[unicode=true, breaklinks=false, pdfborder={0 0 1}, backref=false, colorlinks=true, linkcolor=blue, citecolor=blue]{hyperref}

\usepackage[caption=false]{subfig} %to create subfigures with individual captions
\usepackage{lipsum}

\def\bra#1{\mathinner{\langle{#1}|}}
\def\ket#1{\mathinner{|{#1}\rangle}}
\def\braket#1{\mathinner{\langle{#1}\rangle}}

\def\BraVert{\egroup\,\mid\,\bgroup}

\def\ketbra#1#2{|#1\rangle \!\langle#2|}
\graphicspath{Figures}

\definecolor{Blue}{rgb}{0,0,1}
\definecolor{Red}{rgb}{1,0,0}
\definecolor{Green}{rgb}{0,1,0}
\definecolor{darkgreen}{rgb}{0,.7,0}
\definecolor{Purp}{rgb}{.2,0,.2}
\definecolor{white}{rgb}{1,1,1}

\newcommand{\beq}{\begin{equation}}
\newcommand{\eeq}{\end{equation}}

\DeclarePairedDelimiter\floor{\lfloor}{\rfloor}

\begin{document}
\title{Experimental quantum memristor}
\author{Michele Spagnolo}
\author{Joshua Morris}
\affiliation{Vienna Center for Quantum Science and Technology (VCQ), Faculty of Physics, University of Vienna, 1090 Vienna, Austria}

\author{Simone Piacentini}
\affiliation{Dipartimento di Fisica, Politecnico di Milano, piazza L. Da Vinci 32, 20133 Milano, Italy}
\affiliation{Istituto di Fotonica e Nanotecnologie, Consiglio Nazionale delle Ricerche (IFN-CNR), piazza L. Da Vinci 32, 20133 Milano, Italy}

\author{Michael Antesberger}
\affiliation{Vienna Center for Quantum Science and Technology (VCQ), Faculty of Physics, University of Vienna, 1090 Vienna, Austria}
\author{Francesco Massa}
\affiliation{Vienna Center for Quantum Science and Technology (VCQ), Faculty of Physics, University of Vienna, 1090 Vienna, Austria}

\author{Andrea Crespi}
\affiliation{Dipartimento di Fisica, Politecnico di Milano, piazza L. Da Vinci 32, 20133 Milano, Italy}
\affiliation{Istituto di Fotonica e Nanotecnologie, Consiglio Nazionale delle Ricerche (IFN-CNR), piazza L. Da Vinci 32, 20133 Milano, Italy}

\author{Francesco Ceccarelli}
\affiliation{Istituto di Fotonica e Nanotecnologie, Consiglio Nazionale delle Ricerche (IFN-CNR), piazza L. Da Vinci 32, 20133 Milano, Italy}
\affiliation{Dipartimento di Fisica, Politecnico di Milano, piazza L. Da Vinci 32, 20133 Milano, Italy}

\author{Roberto Osellame}
\affiliation{Istituto di Fotonica e Nanotecnologie, Consiglio Nazionale delle Ricerche (IFN-CNR), piazza L. Da Vinci 32, 20133 Milano, Italy}
\affiliation{Dipartimento di Fisica, Politecnico di Milano, piazza L. Da Vinci 32, 20133 Milano, Italy}

\author{Philip Walther}
\affiliation{Vienna Center for Quantum Science and Technology (VCQ), Faculty of Physics, University of Vienna, 1090 Vienna, Austria}
\affiliation{Christian Doppler Laboratory for Photonic Quantum Computer, Faculty of Physics,  University of Vienna, 1090 Vienna, Austria}

%\date{\today}

\begin{abstract}
    Quantum computer technology harnesses the features of quantum physics for revolutionizing information processing and computing \cite{Gyongyosi2019}. As such, quantum computers use physical quantum gates that process information unitarily, even though the final computing steps might be measurement-based or non-unitary \cite{Raussendorf2001,Walther2005}. The applications of quantum computers cover diverse areas, reaching from well-known quantum algorithms \cite{Grover1997,Shor1999} to quantum machine learning \cite{Biamonte2017,Dunjko2018} and quantum neural networks \cite{Steinbrecher2019}. The last of these is of particular interest by belonging to the promising field of artificial intelligence. However, quantum neural networks are technologically challenging as the underlying computation requires non-unitary operations for mimicking the behavior of neurons \cite{Schuld2014}. A landmark development for classical neural networks was the realization of memory-resistors, or “memristors” \cite{Chua1971,Strukov2008}. These are passive circuit elements that keep a memory of their past states in the form of a resistive hysteresis and thus provide access to nonlinear gate operations. The quest for realising a quantum memristor led to a few proposals \cite{Pfeiffer2016,Salmilehto2017,Sanz2018}, all of which face limited technological practicality. Here we introduce and experimentally demonstrate a novel quantum-optical memristor that is based on integrated photonics and acts on single photons. We characterize its memristive behavior and underline the practical potential of our device by numerically simulating instances of quantum reservoir computing \cite{Nakajima2017,Paterek2019, Nakajima2019}, where we predict an advantage in the use of our quantum memristor over classical architectures. Given recent progress in the realization of photonic circuits for neural networks applications \cite{Steinbrecher2019,Hamerly2019,Goi2020}, our device could become a building block of immediate and near-term quantum neuromorphic architectures.
\end{abstract}

\maketitle
%%%%%%%%%%%%%%%%%%%%%%%%%%%%%%%%%%
%%%%%%%%%%%%%%%%%%%%%%%%%%%%%%%%%%
\section*{Introduction}
In the past few decades, the field of computer science has witnessed two fundamental paradigm shifts. The first relates to artificial neural networks \cite{Anthony2009}. These structures have proven very effective in many relevant tasks, such as speech recognition  \cite{Chan2016}  or image  classification  \cite{Du2017} and nowadays form the core of the most advanced artificial intelligence algorithms  \cite{Nilsson2014}. The second is quantum computation, which harnesses uniquely quantum features such as superposition and entanglement \cite{Gyongyosi2019} to provide dramatic advantages for classically intractable problems \cite{Grover1997,Shor1999,Arute2019}, and enhanced data security for networks and data clouds \cite{Barz2012,Roehsner2018,Tham2020}.

Combining the advantages of neural networks with those of quantum physics may open groundbreaking technological outcomes, and has already attracted much theoretical investigation  \cite{Steinbrecher2019}. The main challenge of this approach lies in the contrast between the intrinsically nonlinear and dissipative nature of neural networks and the linear, unitary dynamics of quantum mechanics. Combining these two contradictory concepts for obtaining an overall advantage has thus far proven to be particularly challenging \cite{Schuld2014}.

One of the fundamental components of biologically inspired neural networks is the memory-resistor, or memristor. Such a device was postulated in 1971 by Leon Chua  \cite{Chua1971} and physically demonstrated for the first time by Struckov et al. in 2008  \cite{Strukov2008}. Since then, various memristive systems were demonstrated on semiconductor platforms \cite{mazumder2012}, including metal-oxide quantum dots \cite{Li2017}.
Remarkably, it was quickly recognised that the particular nonlinear, hysteretic behaviour of the memristor is similar to that of neural synapses  \cite{Linares-Barranco2009}, which opened up new experimental concepts and promising  architectures for neural networks and neuromorphic computing \cite{Jo2010,Yu2011,Moon2019,Yao2020}. 
It is then natural to wonder whether a quantum version of a memristor may be experimentally realized and may provide the necessary features that are required for a large-scale quantum neural network. In this work, we suggest this is the case, as we experimentally demonstrate a feasible architecture for quantum memristors in the photonic domain, and underline its advantages by introducing applications that would benefit significantly from using such quantum devices. 

%\com{To be honest, I feel this last sentence gives a good enough end to the paragraph. The more I read it, the more I think the next sentence is quite unnecessary.} \rem{Inspired by a recent proposal, we introduce what may be seen as a quantum photonics equivalent of Struckov’s memristor: namely, a quantum device that provides a memristive behavior when processing quantum information. Remarkably, this quantum memristor can be realized without quantum memory and with an architecture that allows for preserving a sufficient degree of quantum coherence for subsequent quantum information processing. We design and realise a novel intergrated photonic circuit, which was fabricated via laser-writing technology, for the first demonstration of a quantum memristor. For emphasizing the immediate relevance of quantum memristors, we consider the application of our device to traditionally hard learning tasks that reside in the classical domain (image classification) and in the quantum domain (entanglement witness). In both cases, our numerical evidence suggests a significant advantage of architectures that utilize quantum memristors with respect to conventional strategies.}

%%%%%%%%%%%%%%%%%%%%%%%%%%%%%%%%%%
%%%%%%%%%%%%%%%%%%%%%%%%%%%%%%%%%%
\section*{From classical to quantum memristors}
\begin{figure}
    \centering
    \includegraphics[width=\columnwidth]{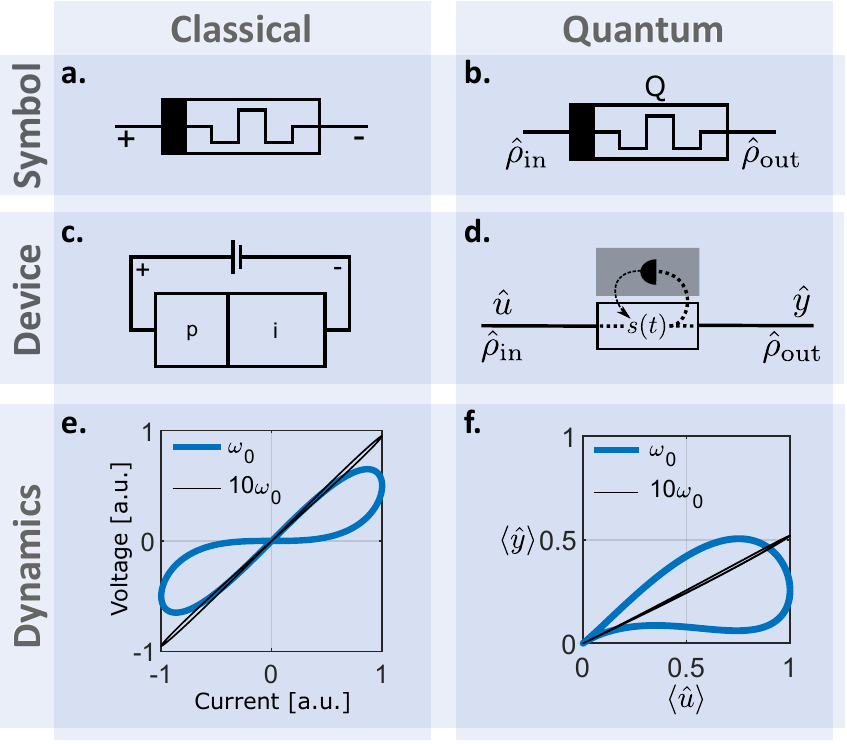}
    \caption{\textbf{Comparison of classical and quantum memristors.} \textbf{(a)} Circuit symbol for classical memristor and \textbf{(b)} quantum memristor. \textbf{(c)} First physical model of a classical electronic memristor based on a junction between doped semiconductor \emph{p} and instrinsic semiconductor \emph{i}. \textbf{(d)} General concept of a quantum memristor: a device that acts on quantum states and whose state variable $s(t)$ is coupled to the environment by a measurement process. The coupling must be engineered so that quantum coherence is sufficiently preserved from the input state $\hat\rho_\text{in}$ to the output state $\hat\rho_\text{out}$. \textbf{(e)} Theoretical dynamics of the classical electronic memristor, showing the signature hysteresis loop pinched at the origin for a given frequency $\omega_0$. Approaching the high-frequency limit, e.g., $10\,\omega_0$, the curve approximates a line. \textbf{(f)} In a quantum memristor, the expectation values of the quantum input observable $\braket{\hat{u}}$ and output observable $\braket{\hat{y}}$ obey the form of Eqs. \eqref{eq_mem1},\eqref{eq_mem2}, thus originating a hysteresis loop pinched at the origin that approximates a line at high frequencies.}
    \label{fig_memristor}
\end{figure}

The memristor was postulated as the fourth fundamental passive circuit element (the other three being resistor, capacitor, and inductor). The fundamental property of such a device is that it retains a memory of its past states in the form of a resistive hysteresis. Chua later introduced the more general concept of memristive devices  \cite{Chua1976}, which are defined by the following coupled equations
\begin{align}
	 y &= f(s,u,t)\, u,\label{eq_mem1} \\
	 \dot{s} &= g(s,u,t), \label{eq_mem2}
\end{align}
where $u$ and $y$ denote input and output variables respectively, and $s$ denotes a state variable, all of which are implicitly assumed to depend on time $t$. When $f(\cdot)$ and $g(\cdot)$ are linear functions of $u$, the input--output relations of these devices show a characteristic figure-of-eight hysteresis loop pinched at the origin that approximates a line when driven at high frequencies. This was first demonstrated for the electronic memristor \cite{Strukov2008}, where $u$, $y$ are current and voltage and $f(\cdot)$ is a generalised resistance.

A quantum memristor must be able to go beyond the classical counterpart by enabling the same behaviour in addition to preserving quantum coherence when processing information encoded in quantum states. These are demanding properties, which previous proposals satisfied only partially \cite{Pfeiffer2016,Salmilehto2017}. Dependent on the choice of input and output variables, a quantum memristor must provide the following features:
\renewcommand{\labelenumi}{(\alph{enumi})}
\begin{enumerate}
    \item Memristive behaviour in the classical limit, i.e. showing the dynamics of Eqs. \eqref{eq_mem1},\eqref{eq_mem2} when the expectation values of the quantum observables are considered.
    \item Quantum coherent processing, i.e., the ability to coherently map a quantum input state onto an output state.
\end{enumerate}
These two requirements are typically mutually exclusive, which poses severe technological challenges. The only circumstance for a quantum photonic device to produce the memory behaviour required in (a) is via interaction with an environment through some form of measurement process. In practice this is always associated with some level of decoherence, thus negating point (b) and rendering this device no different than a classical memristor. For overcoming this contradiction it is necessary to engineer an open quantum system such that features (a) and (b) can coexist: the interaction with the environment must be strong enough for introducing an effective nonlinearity, but at the same time weak enough to sufficiently preserve quantum coherence. The main features of classical and quantum memristor are outlined and compared in Fig. \ref{fig_memristor}.

%%%%%%%%%%%%%%%%%%%%%%%%%%%%%%%%%%
%%%%%%%%%%%%%%%%%%%%%%%%%%%%%%%%%%
\section*{A photonic quantum memristor}
The possibility of realising a quantum memristor in the photonic domain was first pointed out in \cite{Sanz2018}, but the scheme suffered from conceptual and technical drawbacks that severely hindered practical implementations. Here we go beyond the original proposal by introducing a substantially improved scheme suitable for realisation in integrated optics. A detailed comparison of our scheme with respect to the original one is provided in Appendix \ref{sec_comparison}.

\begin{figure}
    \centering
    \includegraphics[width=\columnwidth]{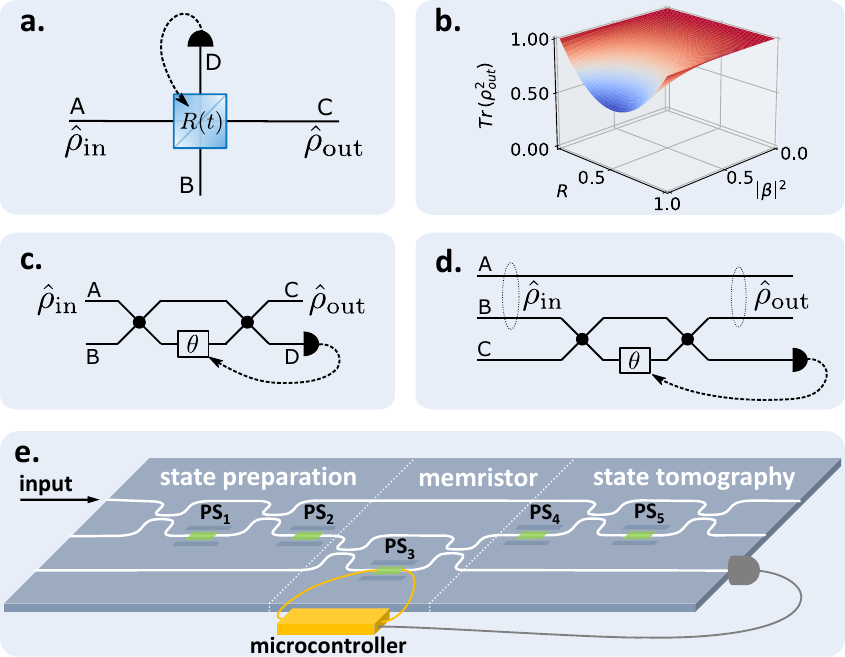}
    \caption{\textbf{Photonic quantum memristor scheme.} \textbf{(a)} The basic concept: a tunable beam splitter with active feedback. Mode $A$ is used as input port, whereas modes $C$ and $D$ are used as output ports. The reflectivity $R(t)$ is updated based on the measurement at mode $D$. \textbf{(b)} The purity $\Tr({\rho_\text{out}}^2)$ of the output state $\rho_\text{out}$ as a function of the refectivity $R$ and the input variable $|\beta|^2$. The state if fully mixed only for $|\beta|^2=1, R=0.5$. \textbf{(c)} The equivalent of a tunable beam splitter in integrated optics: a Mach-Zehnder interferometer, whose reflectivity (i.e., the probability of photons crossing from mode $A$ to mode $D$) is set by a phase shifter in one of the arms. \textbf{(d)} The dual-rail equivalent of the scheme, wherein the input state is encoded as a single photon in a superposition of the two upper modes, one of which goes into the Mach-Zehnder while the other goes directly to the output (see Appendix \ref{sec_chipdesign} for a detailed explanation). \textbf{(e)} The integrated photonics quantum memristor processor, realised by direct laser writing on a glass substrate (see Appendix \ref{sec_FLM}). The chip includes a state-preparation and a state-tomography stage before and after the quantum memristor, respectively. Single-mode fibers are glued to the upper input mode and to the three output modes. The reflectivity of the quantum memristor is externally set by a microcontroller.}
    \label{fig_setup}
\end{figure}

To illustrate the basic principle, let us consider the beam splitter represented in Fig.  \ref{fig_setup}a, whose reflectivity $R(t)$ is tunable and dynamically controlled by an active feedback based on single-photon detection at the output mode $D$. When a quantum state with photon-number expectation value $\braket{n_\text{in}(t)}$ is sent to the input mode $A$ at time $t$, the expectation value $\braket{n_\text{out}(t)}$ at mode $C$ is
\begin{equation}
\braket{n_\text{out}(t)} = \big[1-R(t)\big]\braket{n_\text{in}(t)}.
\label{eq_mem3} 
\end{equation}
The temporal dynamics of the device is determined by the choice of the feedback, i.e., the update rule for $R(t)$. Assuming that $\braket{n_\text{in}(t)}$ takes values between zero and $\braket{n}_\text{max}$, we choose the following relation: 
\begin{equation}
\dot{R}(t) = \braket{n_\text{in}(t)} - 0.5\braket{n}_\text{max}.
\label{eq_mem4} 
\end{equation}
One can see that Eqs. \eqref{eq_mem3},\eqref{eq_mem4} satisfy the form required by Eqs.  \eqref{eq_mem1},\eqref{eq_mem2}, and therefore define a memristive device with $R(t)$ as the state variable. In fact, we show in Appendix \ref{sec_photmem} that Eq.  \eqref{eq_mem3} has a close formal analogy with Struckov's original memristor  \cite{Strukov2008}, which inspired the choice of Eq. \eqref{eq_mem4}. Note that these two equations apply to any input state, which may be also classical light, and thus they do not define a quantum memristor \emph{per se}. 

However, consider now an input state $\ket{\psi_\text{in} (t)}$ in the quantum superposition
\beq 
\label{eq_psi_in}
\ket{\psi_\text{in} (t)} = \alpha (t)\ket{0}_\text{A} + \beta (t)\ket{1}_\text{A},
\eeq 
where $|\alpha (t)|^2+|\beta (t)|^2=1$, and $\ket{0}_\text{A}$ and $\ket{1}_\text{A}$ represent vacuum and a single-photon state in mode $A$, respectively. In this single-photon case $\braket{n_\text{in}(t)}=|\beta(t)|^2$ and  $\braket{n}_\text{max}=1$. If the photon is detected in $D$, the output at mode $C$ is just the vacuum state $\ket{0}_\text{C}$. However, when the photon is \emph{not} detected in $D$, then the output state $\ket{\psi_\text{out,C}(t)}$ at mode $C$ is projected onto
%\begin{linenomath*}
\beq 
\label{eq_psiout0}
\ket{\psi_\text{out,C}(t)}= \frac{\alpha(t)}{\sqrt{N}}\ket{0}_\text{C}+\frac{\beta(t)\sqrt{1-R(t)}}{\sqrt{N}}\ket{1}_\text{C}\,,
\eeq 
%\end{linenomath*}
($N$ being the normalisation factor) which is still a quantum superposition, thus proving that this device provides a genuine quantum behaviour. Intuitively, assuming the user only has access to output $C$, the overall output state is given by the statistical mixture of both cases, weighted by their respective probability: 
\begin{multline}
\label{eq_rhoout}
\rho_\text{out,C}(t) = |\beta(t)|^2R(t)\,\ketbra{0}{0}_\text{C} + \\ + \big[1-|\beta(t)|^2R(t)\big]\,\ketbra{\psi_\text{out,C}}{\psi_\text{out,C}}
\end{multline}
(a formal derivation is provided in Appendix \ref{sec_densmat}). The purity of this state can be calculated as 
\beq
\label{eq_purity}
\Tr({\rho_\text{out}}^2(t)) = 1-2|\beta(t)|^4R(t)(1-R(t)),
\eeq

and is shown as a function of the reflectivity $R$ and the input variable $|\beta|^2$ in Fig. \ref{fig_setup}b. The fact that the state is not fully mixed (except for the case of $|\beta|^2 = 1, R=0.5$) shows that the device is capable of preserving some measure of quantum coherence, thus satisfying the requirements of a quantum memristor.

%\rem{These features became possible by introducing conceptual and technological improvements that overcome the limitations of previous proposals  \cite{Sanz2018}. By establishing the photon number as input variable and single-photon detection as feedback we obtain two important advantages: firstly, the protocol is more experimentally feasible by not requiring quadrature measurements, and secondly by providing a correct memristive behaviour due to hysteresis curves that are pinched at the origin. Remarkably, the pinched shape is not an incidental feature of the memristor but a fundamental one, as it follows directly from Eq.  \eqref{eq_mem1}.}

%%%%%%%%%%%%%%%%%%%%%%%%%%%%%%%%%%
%%%%%%%%%%%%%%%%%%%%%%%%%%%%%%%%%%
\section*{Implementation and results}
\begin{figure*}
    \centering
    \includegraphics[width=\textwidth]{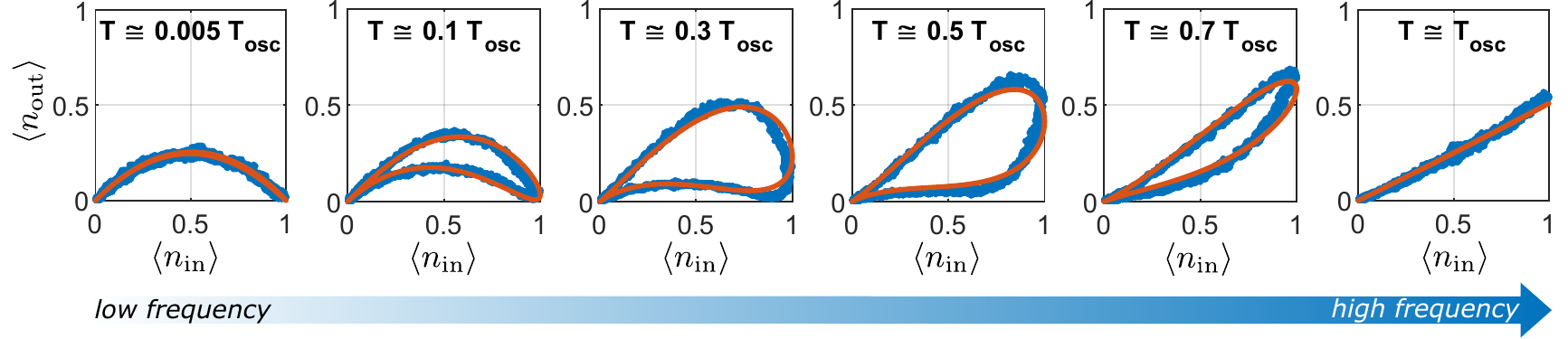}
    \caption{\textbf{Characterisation of the photonic quantum memristor.} Experimental results (blue lines) and simulated dynamics (red lines) for different frequency regimes. The oscillation period is kept constant at $T_\text{osc}=10$s, and the integration time $T$ is varied in the range of one period. Since the high frequency limit is, in this case, the same as $T=T_\text{osc}$, this provides a full characterisation of the dynamic response of the device (see Appendix \ref{sec_feedback} for further details). The experimental data is in perfect agreement with the simulated dynamics. Specifically, the low frequency limit is $\braket{n_\text{out}}_\text{LF} = \braket{n_\text{in}} - \braket{n_\text{in}}^2$, while the high frequency limit is $\braket{n_\text{out}}_\text{HF} = 0.5\braket{n_\text{in}}$. This is also in perfect agreement with the original definition of memristive device \cite{Chua1976}.}
    \label{fig_result}
\end{figure*}

The input state proposed in Eq. \eqref{eq_psi_in} encodes a qubit as a superposition of two energy levels. Despite offering an intuitive picture and a ready comparison across different quantum platforms, this type of encoding (also known as single-rail encoding) is highly impractical in linear optics \cite{Bimbard2010, Filippov2011}. A more natural approach in quantum photonics is path-encoding (also known as dual-rail), wherein the qubit is represented by a single photon being present in either of two spatial modes. In Appendix \ref{sec_chipdesign} we show how the single-rail protocol we presented in the previous paragraph has a straightforward dual-rail equivalent. Practically, one just needs to introduce an additional spatial mode which does not go through the beam splitter. Figure \ref{fig_setup}(c-e) summarises the steps from the basic concept to the final integrated photonic processor. 

The photonic quantum memristor processor is realised by femtosecond laser micromachining \cite{Corrielli2018,Marshall2009}. All sections are fully configurable by means of thermal phase shifters \cite{Ceccarelli2019,Dyakonov2018} featuring novel thermal isolation structures that strongly reduce the power consumption and thermal crosstalk \cite{Ceccarelli2020}. Fabrication details are reported in Appendix \ref{sec_FLM}.

The reflectivity of the on-chip quantum memristor stage is externally set by a microcontroller, which approximates the solution of Eq. \eqref{eq_mem4} by performing a time-window integration of the form
\beq 
\label{eq_integration2}
R(t) = 0.5 + \frac{1}{T}\int_{t-T}^t{(\braket{n_\text{in}(\tau)} - 0.5)}\,d\tau,
\eeq
where $T$ is the width of the integration window (derivation is provided in Appendix  \ref{sec_feedback}). A challenge in implementing this operation is that the measurement of the expectation value $\braket{n_\text{in}}$ requires itself some form of windowed integration of the input signal. Such window needs to be large enough to collect meaningful photon statistics, while still being much smaller than $T$ so that, on the time scale of the memristor, $\braket{n_\text{in}}$ can be considered to be an instantaneous quantity. Our solution, along with a full description of the experimental setup, is detailed in Appendix  \ref{sec_setup}, where we show that $\braket{n_\text{in}}$ is estimated on a time window of approximately 100 ms, corresponding to a few hundred photon counts on average.

A stream of single photons is coupled via single-mode fiber to the upper mode of the chip (see Fig. \ref{fig_setup}e) and, using the integrated state-preparation stage, the input number of photons to the memristor is varied in time as 
\beq
\braket{n_\text{in}(t)} = |\beta^2(t)| = \sin^2(\pi/T_\text{osc}\,t),
\eeq
where $T_\text{osc}$ is the oscillation period. The dynamics of the device is determined by the ratio $T/T_\text{osc}$. We refer to high frequency regime when the input oscillates many times within an integration window, i.e. $T\gg T_\text{osc}$, and conversely to low frequency regime when $T\ll T_\text{osc}$.

An upper bound to $f_\text{osc}=1/T_\text{osc}$ is given by the response of the thermal phase shifters of the chip, which can be modelled as low-pass filters with a cutoff frequency $f_\text{cut} \simeq 5$ Hz. Notably, we observed that when $f_\text{osc}$ approaches this frequency range, it causes additional memristive behaviour, which we describe in Appendix \ref{sec_lpfmemristor}. In Figure  \ref{fig_result} we report instead our results when keeping a constant $f_\text{osc} = 0.1$ Hz (well below $f_\text{cut}$) and varying the integration time $T$. The device shows a hysteresis figure pinched at the origin which reduces to a linear relation at high frequencies, and to a nonlinear one at low frequencies. This is precisely Chua's definition of a memristive device  \cite{Chua1976}.

For further demonstrating the functionality of the quantum memristor, we have characterised the output state with respect to the input state and reflectivity $R$. As an example, for $|\beta|^2=0.3$ and $R=0.7$, we experimentally reconstruct the density matrix with a fidelity of $F=99.7$\% to the theoretical one. The purity of the state is measured to be $\Tr({\rho_\text{out,EXP}}^2) = 0.66$, which matches the theoretical value of $\Tr({\rho_\text{out,THEO}}^2) = 0.67$, showing that our quantum memristor does not significantly introduce additional decoherence. In Appendix \ref{sec_expdensmat} the details of the reconstruction are shown together with 16 output states with an average fidelity of $F=98.8$\%.

%%%%%%%%%%%%%%%%%%%%%%%%%%%%%%%%%%
%%%%%%%%%%%%%%%%%%%%%%%%%%%%%%%%%%
\section*{A memristor-based quantum reservoir computer}
\begin{figure}
    \centering
    \includegraphics[width=\columnwidth]{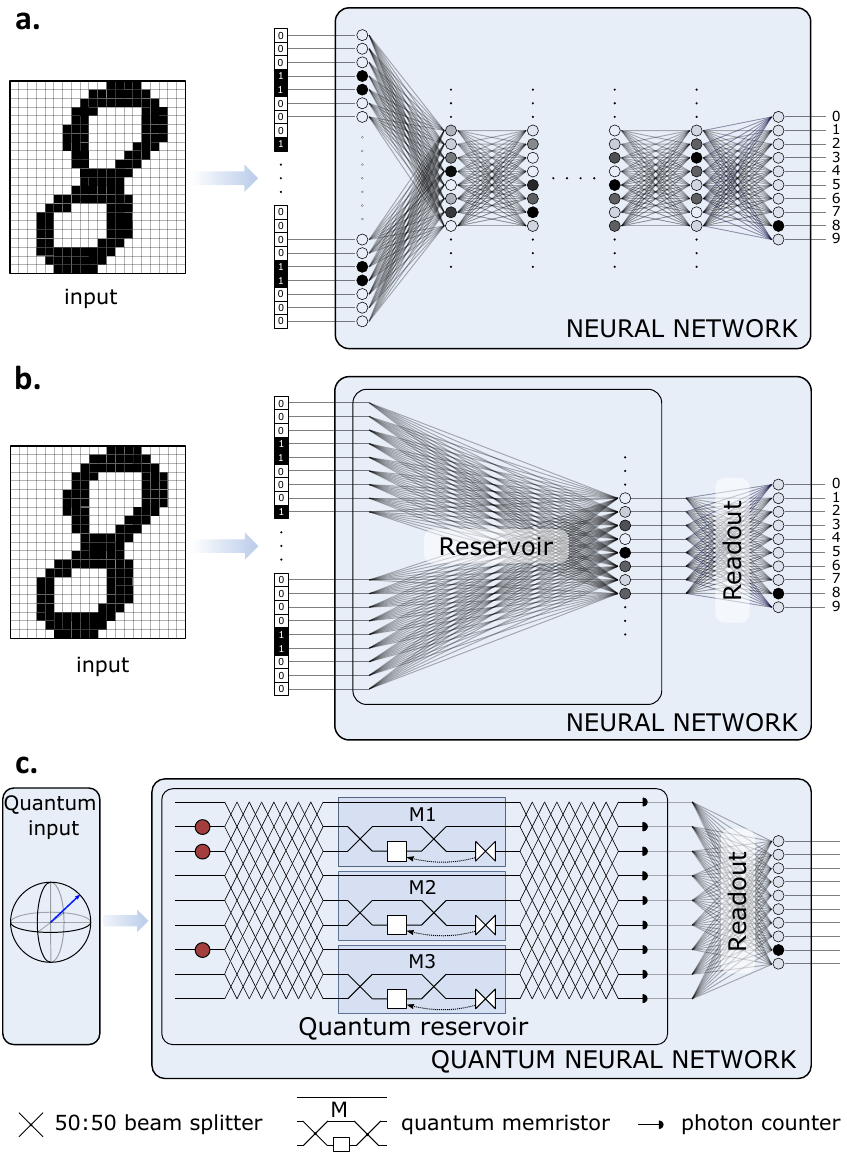}
    \caption{\textbf{Quantum reservoir computing.} \textbf{(a)} A classical neural approach to a classification problem. The input information (e.g. the pixels of an image) is fed to the first layer. The network can be trained to "switch-on" the neuron in the last layer corresponding to the correct class. \textbf{(b)} Reservoir computing. The input information is mapped to a nonlinear, high-dimensional space, whose output is interpreted by a simple linear readout newtork. Only the readout network is trained, which requires minimal resources. \textbf{(c)} A quantum reservoir computer based on quantum memristors. The input information is encoded on quantum states of three photons in nine optical modes. A fixed matrix of beam splitters with random reflectivity distributes the information across all optical modes, which are fed to three quantum memristors, whose outputs are scrambled again before reaching photon counters. Reinjection of the photons is assumed if one is measured at the feedback port of a memristor. The output pattern is interpreted by a trainable linear readout function.}
    \label{fig_newnet}
\end{figure}

Neural networks are known to be very effective in computational tasks where a small amount of information (e.g., whether an image represents a cat or a dog) needs to be extracted from high-dimensional data (e.g. an image matrix of thousands of pixels). Typical neural approaches to this problems involve densely-connected, multi-layer structures like the one schematised in Fig. \!\ref{fig_newnet}a. While having been proven extremely effective, training these networks requires an iterative optimisation of thousands, sometimes millions of parameters, which in turn requires very large amounts of high quality training data and computational time. This issue represents the main limiting factor for the scalability of these architectures.

Reservoir computing \cite{Herbert2001, Maass2002} addresses this challenge by having the input data processed through a fixed nonlinear high-dimensional system (a reservoir). This reservoir maps the data such that the output only requires an elementary readout network for being interpreted, e.g. a linear classifier (Fig. \ref{fig_newnet}b). One key advantage of this approach is that only the readout network needs to be trained, which requires minimal resources in both time and data. Secondly, reservoirs can be implemented on physical systems rather than computer models, which promises even further speedups \cite{tanaka2019}. Classical physical reservoirs have been demonstrated on a variety of platform, including classical memristors \cite{Du2017,Moon2019} and classical optics \cite{duport2012, vandoorne2014, rafayelyan2020}. Considerable interest has been recently devoted to quantum reservoirs \cite{Nakajima2017,Paterek2019, Nakajima2019, negoro2018, mujal2021}. Here we propose and numerically evaluate a quantum photonic reservoir based on quantum memristors.

Figure \ref{fig_newnet}c schematises the working principle of the quantum reservoir computer. In this simulated example the input information is encoded as quantum states represented by three photons that can occupy nine different optical modes. A fixed matrix of beam splitters with randomly assigned reflectivity scrambles the information across all optical modes, which is then fed into the input ports of three quantum memristors. The outputs of the quantum memristors are scrambled again before reaching an array of photon counters. Note that the system is inherently resilient to photon losses, as the detectors always herald the three-fold events. In the end, this detected output signal is then fed into the readout network. It has been shown \cite{mujal2021} that reservoir computing provides excellent performances when having access to i) high dimensionality, ii) non-linearity resources, and iii) short-term memory. Here we propose a quantum reservoir that combines passive optical networks with our demonstrated quantum memristors. The photonic network gives access to a large Hilbert space that grows exponentially with the size of the quantum system. In contrast, the nonlinearity and short-term memory are provided by the quantum memristors. This is a key difference with respect to the scheme of \cite{Nakajima2017}, where the nonlinearity and memory arise from the dynamics of the ensemble of solid-state qubits.

\paragraph*{Image classification by sequential data analysis.} Reservoir computing is naturally suited for interpreting time-dependent data. Image classification, although usually regarded as a static task, can be reframed as a time-dependent task when considering images as pixels whose arrangement is defined by an ordered sequence of columns. Such an approach provides the advantage that the instantaneous input dimension is greatly reduced, as it only needs to encode one column at a time, rather than the whole pixel matrix. A second, more practical advantage, is that very high quality image databases are available. We consider here a subset of the MNIST handwritten digit database \cite{lecun2010mnist} representing digits "0", "3" and "8" (chosen for their column-wise similarity). Each image is cropped to 18x12 pixels, and the columns are encoded one at a time into the quantum reservoir via simple amplitude encoding scheme (see Appendix \ref{sec_QRC}). At each step, the state of the quantum memristors is updated via a discrete time equivalent of Eq. \eqref{eq_integration2}. The output corresponding to the last column is finally interpreted by the linear readout network, which is composed of approximately 1600 tunable parameters. After training on one thousand different images over 15 epochs, we achieve a classification accuracy of 95\% on a never before seen test set of one thousand images split evenly across the chosen digits. 

Remarkably, our analysis shows that high accuracy was achieved on this three-digit classification task by using only an extremely small training set of just one thousand images, using a very small physical reservoir containing only three quantum memristors, and a very small readout network. Although comparing the performances of neural networks is challenging as they tend to be case-specific, reported classical schemes require more resources for similar tasks. The authors of \cite{Du2017}, who implemented a similar scheme, reported a simulated 91\% accuracy on the ten-digit classification using fourteen thousand training images and a reservoir containing eighty-eight classical memristors. In \cite{jalalvand2015} an accuracy of 92\% was reached with three layered reservoirs, sixty thousand training images and approximately five hundred thousand tunable parameters. It thus seems plausible to conclude that our scheme is highly resource-efficient compared to existing ones. Whether such efficiency reflects a genuine quantum advantage associated to the quantum reservoir remains to be discussed. Although numerical evidence has been often reported, a full proof of quantum advantage is still an active field of research. 

\begin{figure}
    \centering
    \includegraphics[width=\columnwidth]{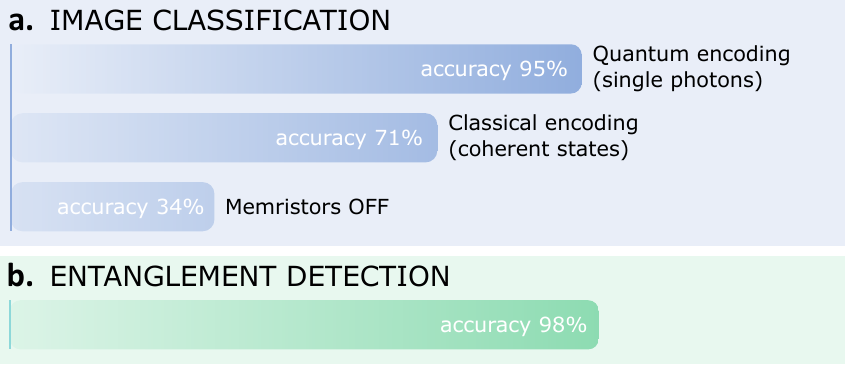}
    \caption{\textbf{Performance of quantum reservoir computing using quantum memristors.} A quantum reservoir computer is simulated that is composed of of three photons, nine modes, three quantum memristors and a final read-out network. \textbf{(a)} Classical task. For the image classification of three different digits the simulations show the best accuracy when the input is encoded as quantum state using single photons. Encoding with coherent light significantly degrades the performance, indicating superior performance when processing quantum data. Switching off the quantum memristors in the reservoir reduces the accuracy to essentially random guessing. \textbf{(b)} Quantum task. For the discrimination of quantum input states, whether to be separable or maximally entangled, the same quantum reservoir computer allows to distinguish these states with high accuracy.}
    \label{fig_accuracies}
\end{figure}

Therefore for obtaining insights into the quantum advantage of quantum memristors, we have compared the performance of our quantum reservoir computation to the on obtained when using only classical information as input (Fig. \ref{fig_accuracies}a). This was achieved by encoding the input information with coherent classical light, rather than single photons, and by keeping all other conditions the same. The resulting accuracy for distinguishing the three digits dropped to approximately 71\%, which indicates a superior performance of the quantum case. Also, when switching off the feedback loop of the quantum memristor, which eliminates both the nonlinearity and the memory from the reservoir, then the performance drops to 34\%, which is essentially random guessing for a three label classification task.   

\paragraph*{Entanglement detection.} Naturally, a quantum reservoir is also suited for quantum tasks that are inaccessible with classical resources. For demonstrating this potential, we took the same quantum reservoir computer as for the classical image classification. As exemplary task for quantum applications we analyzed the capability of detecting quantum entanglement as a two-way discrimination problem between separable and maximally-entangled quantum states. One-hundred copies of each state are fed to the quantum reservoir sequentially, and the state of the quantum memristors is updated based on the measurement statistics collected from this sequence. For this specific task the quantum memristors’ nonlinearity rather than its memory is exploited for increasing the complexity of the map performed by the quantum reservoir. By training on a set of just 1000 randomly generated pure states we obtain a discrimination accuracy of 98\% (Fig. \ref{fig_accuracies}b), which indicates that the network has effectively learned to generate a relatively high-performing entanglement detection protocol with nil user input.

\section*{CONCLUSIONS}
We have designed an optical memristive element that allows the transmission of coherent quantum information as a superposition of single photons on spatial modes. We have realised the prototype of such a device on a glass-based, laser-written photonic processor and thereby provided the first experimental demonstration of a quantum memristor. We have then designed a memristor-based quantum reservoir computer and tested it numerically on both classical and quantum tasks, achieving significant performances with very limited physical and computational resources and, most importantly, no architectural change from one to the other.

Our demonstrated quantum memristor is feasible in practice and easily scalable to larger architectures using integrated quantum photonics. Additionally, the frequency at which our quantum memristor operates can be easily improved by many orders of magnitudes. For laser-written circuits high frequency operations are readily available at the expense of higher power consumption \cite{Ceccarelli2020},  whereas other photonic platforms routinely enable frequencies even in the GHz regime \cite{reed2010}. For exploiting these frequencies for integrated photonic circuits the photon detection rate must be improved as well. The vast development of quantum photonics technology shows that such performances are in reach by using customized fast detectors and bright single-photon sources using quantum dots \cite{Senellart2017}. 

We emphasise that our results are not restricted to photonic quantum systems, and would be equally applicable to other platforms such as superconducting qubits \cite{Pfeiffer2016,Salmilehto2017}. On the other hand, our photonic implementation offers a particularly simple and robust approach that relies on a mature technological platform, and may provide the missing nonlinear element for recently proposed quantum optical neural networks \cite{Steinbrecher2019}. Given the recent progress in photonic circuits for neuromorphic applications\! \cite{Shen2017}, we envisage our device to play a key role of future photonic quantum neural networks.
%%%%%%%%%%%%%%%%%%%%%%%%%%%%%%%%%%
%%%%%%%%%%%%%%%%%%%%%%%%%%%%%%%%%%
%TC:ignore
\begin{acknowledgements}
The authors are thankful to V. Saggio, T. Strömberg, P. Schiansky, B. Daki\'c, B. Peterson, L. Rozema, G. Zanin, I. A. Calafell, G. Bellomia and the PoliFAB staff (www.polifab.polimi.it) for advice and support. P.W. acknowledges support from the research platform TURIS, the European Commission through UNIQORN (no. 820474), and EPIQUS (no. 899368), from the Austrian Science Fund (FWF) through CoQuS (W1210-4), BeyondC (F7113) and Reseach Group 5 (FG5), from the AFOSR via PhoQuGraph (FA8655-20-1-7030), and from the Austrian Federal Ministry for Digital and Economic Affairs, the National Foundation for Research, Technology and Development and the Christian Doppler Research Association. R.O. acknowledges financial support from the European Research Council (ERC) under the European Union's Horizon 2020 research and innovation programme (project CAPABLE, No. 742745). P.W. and R.O. acknowledge financial support from the European Commission through HiPhoP (no. 731473).
\end{acknowledgements}

Data and code are available upon request.

%%%%%%%%%%%%%%%%%%%%%%%%%%%%%%%%%%
%%%%%%%%%%%%%%%%%%%%%%%%%%%%%%%%%%
%\clearpage
\bibliography{biblio}

%%%%%%%%%%%%%%%%%%%%%%%%%%%%%%%%%%
%%%%%%%%%%%%%%%%%%%%%%%%%%%%%%%%%%
\clearpage
%\begin{appendices}
\section*{Appendix}
\subsection{Output state of the quantum memristor}
\label{sec_densmat}

Let us rewrite for convenience the input state
\beq 
\tag{\ref{eq_psi_in}}
\ket{\psi_\text{in} (t)} = \alpha (t)\ket{0}_\text{A} + \beta (t)\ket{1}_\text{A}.
\eeq 
When going through a beam splitter with reflectivity $R(t)$ the state evolves into:
\begin{multline} 
\label{eq_psi_out}
\ket{\psi_\text{out,CD}(t)} = \alpha(t)\,\ket{0}_\text{C}\ket{0}_\text{D} + \\ +\beta(t)\sqrt{1-R(t)}\,\ket{1}_\text{C}\ket{0}_\text{D} + \\
+ i\beta(t)\sqrt{R(t)}\,\ket{0}_\text{C}\ket{1}_\text{D},
\end{multline} 
which corresponds to the following density matrix (let us omit the time $t$ and the pedices $C$ and $D$ for ease of reading):
\begin{align*} 
&\rho_\text{out,CD} = \ket{\psi_\text{out,CD}}\bra{\psi_\text{out,CD}} = |\alpha|^2\,\ketbra{00}{00} + \\
&+\alpha\beta^*\sqrt{1-R}\,\ketbra{00}{10} - i\alpha\beta^*\sqrt{R}\,\ketbra{00}{01} + \\ &+\alpha^*\beta\sqrt{1-R}\,\ketbra{10}{00} + |\beta|^2(1-R)\,\ketbra{10}{10} + \\ &-i|\beta|^2\sqrt{R(1-R)}\,\ketbra{10}{01} + i\alpha^*\beta\sqrt{R}\,\ketbra{01}{00} + \\ &+i|\beta|^2\sqrt{R(1-R)}\,\ketbra{01}{10} + |\beta|^2R\,\ketbra{01}{01}. 
\end{align*} 
Note that output $D$ is used for the feedback of the quantum memristor, whereas the user has only access to output $C$. The output state related to mode $C$ can be obtained from the previous density matrix by taking the partial trace over $D$, and results in:
\begin{multline} 
\label{eq_densmat_part}
\rho_\text{out,C} = \Tr_\text{D}(\rho_\text{out,CD}) = |\alpha|^2\,\ketbra{0}{0}_\text{C} + \\
+\alpha\beta^*\sqrt{1-R}\,\ketbra{0}{1}_\text{C} + \alpha^*\beta\sqrt{1-R}\,\ketbra{1}{0}_\text{C} + \\
+ |\beta|^2(1-R)\,\ketbra{1}{1}_\text{C} + |\beta|^2R\,\ketbra{0}{0}_\text{C},
\end{multline}
which is equal to the one shown in Eq. \eqref{eq_rhoout}. In the matrix representation for the basis $\ket{0}_\text{C}, \ket{1}_\text{C}$ this state corresponds to:
\begin{equation}
    \label{eq_densmat}
    \rho_\text{out,C} =
    \begin{pmatrix}
    |\alpha|^2+|\beta|^2R & \alpha^*\beta\sqrt{1-R} \\ 
    \alpha\beta^*\sqrt{1-R} & |\beta|^2(1-R)
    \end{pmatrix}
\end{equation}
from which the purity is given by
\beq
\tag{\ref{eq_purity}}
\Tr({\rho_\text{out,C}}^2(t)) = 1-2|\beta|^4R(1-R),
\eeq

\subsection{Converting photon-number encoding to path-encoding.}
\label{sec_chipdesign}

In Appendix \ref{sec_densmat} we have developed the theory of the quantum memristor with an encoding scheme where the qubit is represented by a superposition of Fock states, also known as single-rail encoding. In a path-encoded picture (also known as dual-rail) the qubit is represented instead by a single photon being in a superposition of two spatial modes. The $\ket{0}$ state corresponds to the photon being in one mode, say mode $A$, and the $\ket{1}$ state corresponds to the photon being in the other mode, say mode $B$. In short, the map from number encoding to path encoding is the following:
\begin{align}
    \label{eq_map}
    \ket{0}_\text{A} \quad &\rightarrow \quad \ket{1}_\text{A}\ket{0}_\text{B} \\
    \ket{1}_\text{A} \quad &\rightarrow \quad \ket{0}_\text{A}\ket{1}_\text{B}
\end{align}
so our equivalent input state will be
\beq
\label{eq_psiin_dr}
\ket{\psi_\text{in} (t)} = \alpha (t)\,\ket{1}_\text{A}\ket{0}_\text{B} + \beta (t)\,\ket{0}_\text{A}\ket{1}_\text{B}.
\eeq

By looking at Eq. \eqref{eq_psi_out} it is evident that the $\ket{0}_\text{A}$ state is not affected by the beam splitter, which only acts on the $\ket{1}_\text{A}$ state. As a result, the dual-rail equivalent should be one where mode $A$ goes directly to the output, whereas mode $B$ goes into the tunable beam splitter (or its integrated equivalent, i.e. a tunable Mach-Zehnder). This is precisely the scheme shown in Fig. \ref{fig_setup}d. By evolving the input state through the Mach-Zehnder we obtain
\begin{multline} 
\label{eq_rhoout_abc}
\ket{\psi_\text{out,ABC}(t)} = \alpha(t)\,\ket{1}_\text{A}\ket{0}_\text{B}\ket{0}_\text{C} + \\ +\beta(t)\sqrt{1-R(t)}\,\ket{0}_\text{A}\ket{1}_\text{B}\ket{0}_\text{C} + \\
+ i\beta(t)\sqrt{R(t)}\,\ket{0}_\text{A}\ket{0}_\text{B}\ket{1}_\text{C},
\end{multline}
which is the dual-rail equivalent of Eq. \eqref{eq_psi_out}. The output mode $C$ is used as measurement port for updating the state of the quantum memristor, whereas the user only has access to the output modes $A$ and $B$. In order to obtain the output states of modes $A$ and $B$, one has to write $\ket{\psi_\text{out,ABC}(t)}$ in terms of density matrix and then take the partial trace over $C$. With similar calculations to the ones presented in the previous paragraph, one obtains the following density matrix
\begin{equation}
    \label{eq_densmat_dr}
    \rho_\text{out,AB} =
    \begin{pmatrix}
    |\beta|^2R & 0 & 0 \\
    0 & |\alpha|^2 & \alpha^*\beta\sqrt{1-R} \\ 
    0 & \alpha\beta^*\sqrt{1-R} & |\beta|^2(1-R)
    \end{pmatrix}
\end{equation}
(written in the basis $\ket{00}_\text{AB}, \ket{01}_\text{AB}, \ket{10}_\text{AB}$). We can straightforwardly obtain the purity
\beq
\label{eq_purity_dr}
\Tr({\rho_\text{out,AB}}^2(t)) = 1-2|\beta|^4R(1-R),
\eeq
which is the same as in Eq. \eqref{eq_purity}. As in the single-rail case, we take $\braket{n_\text{in}}$ and $\braket{n_\text{out}}$ to represent the number of photons going in and out of the Mach-Zehnder and this corresponds to $\braket{n_\text{in}(t)} = |\beta(t)|^2$, $\braket{n}_\text{max} = 1$ and $\braket{n_\text{out}(t)} = |\beta(t)|^2\big[1-R(t)\big] = \big[1-R(t)\big] \braket{n_\text{in}(t)}$. This shows for the quantum memristor that the two pictures are perfectly equivalent.

The advantage of having switched to dual-rail encoding is, unlike the single-rail encoding, the straightforward manipulation of the qubit. In fact, any arbitrary state of the form of Eq. \eqref{eq_psiin_dr} can be generated by a Mach-Zehnder interferometer with a tunable phase shifters in one of the output arms. This configuration was used as preparation stage for the quantum memristor, and a similar configuration as a final tomography stage after the quantum memristor for choosing arbitrary measurment basis enabling the reconstruction of the density matrix of the output state (see Fig. \ref{fig_setup}e). This tomography stage was only used for characterization purposes and otherwise set to perform an identity operation.
\subsection{Fabrication of the integrated  photonic quantum processor}
\label{sec_FLM}

The fabrication of the integrated photonic chip is based on the femtosecond laser micromachining process \cite{Ceccarelli2020}. Single-mode optical waveguides, optimised for operation at 1550 nm, are inscribed in a alumino-borosilicate glass (Corning EAGLE XG, 1.1 mm thick) by focusing laser pulses (Yb:KYW cavity-dumped mode-locked laser: 1030 nm wavelength, 300 fs pulse duration, 520 nJ energy per pulse, 1 MHz repetition rate) with a 50$\times$ objective (0.65 NA) equipped with an aberration-correction collar. The entire optical circuit is inscribed at 25 $\mu$m from the bottom surface of the substrate, by translating the substrate at the constant speed of 40 mm/s. In particular, six overlapped laser scans are performed along the desired waveguide path. In order to obtain single-mode operation and reduce the waveguide birefringence \cite{Corrielli2018}, the inscription process is followed by a thermal annealing composed of a fast rising ramp of 12 °C/min up to 750 °C and by two subsequent slow falling ramps of 12 °C/h and 24 °C/h, respectively down to 630 °C and 500 °C. After that, the cooling process is completed with no control on the temperature ramps. At the end of the waveguide fabrication process, the measured insertion loss is 1.2 dB, corresponding to a transmission of 76\%. Each Mach-Zehnder interferometer is composed of two balanced directional couplers (zero interaction length and 7.5 $\mu$m coupling distance), that are connected to the rest of the circuit by S-bend waveguides (40 mm curvature radius) and by straight waveguides (separation $p$ = 127 $\mu$m and length $L$ = 2 mm). 

In order to guarantee maximum efficiency and minimal crosstalk of the phase shifting operation, thermal insulating trenches are ablated at both sides of the optical waveguides that are supposed to be phase-tuned. To fabricate the trenches we used laser pulses (Light Conversion PHAROS: 1030 nm wavelength, 1 ps pulse duration, 1.5 $\mu$J energy per pulse, 20 kHz repetition rate) focused by a 20$\times$ water-immersion objective (0.50 NA) on the bottom surface of the substrate, while the latter is translated at 4 mm/s entirely immersed in distilled water. This fabrication technique is usually referred to as water-assisted laser ablation \cite{Li2013,Murakami2017}. In order to realise a single trench with depth $D_t$ = 300 $\mu$m, width $W_t$ = 97 $\mu$m and length $L_t$ = $L$ = 2 mm, four rectangular glass blocks (depth $D_b$ = $D_t/4$ = 75 $\mu$m) are removed one after the other by ablating only the perimeter of each block and making it detach and fall into the water. In this way, deep trenches are fabricated on the bottom side of the substrate with near-unity yield.

After that, the substrate is flipped and the process continues on the bottom side with the fabrication of the thermal phase shifters \cite{Ceccarelli2019}. Firstly, after a standard piranha cleaning bath, a metal multilayer film, composed of 3 nm of chromium and 100 nm of gold, is deposited on the entire area of the chip by using a magnetron sputtering system. Secondly, a further thermal annealing (rising ramp of 10 °C/min up to 500 °C, followed by 60 min at this temperature and by a cooling process with no thermal actuation) is employed to reach a stable value of the electrical resistivity and to prevent electrical drifts that would impair the stability of the phase shifting operation. Lastly, the thermal phase shifters are patterned by laser pulses (Yb:KYW cavity-dumped mode-locked laser: 1030 nm wavelength, 300 fs pulse duration, 200 nJ energy per pulse, 1 MHz repetition rate) focused on the chip surface with a 10$\times$ objective (0.25 NA). By translating the substrate at 2 mm/s, contact pads and electrodes are isolated by selectively removing the metal. Resistive microheaters having width $W_r$ = $p - W_t$ = 30 $\mu$m and length $L_r$ = $L$ = 2 mm are instead isolated by the presence of the trenches. The average electrical resistance of the microheaters is 38 $\Omega$, while the resulting electrical power needed to induce a $2\pi$ phase shift is as low as 55 mW.

In the end, the photonic chip is mounted on an aluminium heat sink, wire-bonded to a printed circuit board and pigtailed to both input and output single-mode optical fibers. After the pigtailing process, the total insertion loss from input to output fibers is 2 dB, corresponding to a transmission of 63\%.

\subsection{Implementation of the feedback law for the quantum memristor}
\label{sec_feedback}

The general solution of Eq. \eqref{eq_mem4} is
\beq 
\label{eq_integration1}
R(t) = c + \int_0^t{(\braket{n_\text{in}(\tau)} - 0.5)}\,d\tau.
\eeq
where we assumed $\braket{n}_\text{max}=1$ and $c$ an arbitrary constant. In our case we want to set $c = 0.5$ in order to restore the baseline of $R$ to zero. 

Even more importantly, we must make sure that in all cases $R(t)$ remains bound in the interval $[0,1]$. This means that a saturation mechanism must be introduced so that the integral does not diverge when $\braket{n_\text{in}(t)}$ is constant or slowly varying. The most physically meaningful way to do so is to integrate over a sliding time window of width $T$, which can be written as
\beq 
\tag{\ref{eq_integration2}}
R(t) = 0.5 + \frac{1}{T}\int_{t-T}^t{(\braket{n_\text{in}(\tau)} - 0.5)}\,d\tau.
\eeq
In other words, at every given time $t$ the memristor "forgets" what happened before time $t-T$. From a physical point of view this is meaningful because no real device can keep memory over an infinitely long time.

The resulting dynamics of our device can be tested by assuming, for example, a periodic input of the form $\braket{n_\text{in}(t)} = \sin^2 (\pi/T_\text{osc}\,t)$. Depending on the relation between the oscillation period of the input, $T_\text{osc}$, and the integration time, $T$, two limiting regimes can already be identified. When $T_\text{osc} \gg T$, i.e. the input is (almost) constant, Eq. \eqref{eq_integration2} reduces to $R(t) = \braket{n_\text{in}}$, which inserted into Eq. \eqref{eq_mem3} gives
\beq
\label{eq_lowfreq}
\braket{n_\text{out}}_\text{LF} = \braket{n_\text{in}} - \braket{n_\text{in}}^2.
\eeq
In contrast, when the input oscillates very quickly such that $T_\text{osc} \ll T$, the integral tends to zero, so that $R(t) = 0.5$ and consequently
\beq
\label{eq_hifreq}
\braket{n_\text{out}}_\text{HF} = 0.5\braket{n_\text{in}}. 
\eeq
Note that when $T=T_\text{osc}$ the integral is zero, as we are integrating over a full period. Hence, integrating over more than one period yields redundant results. For this reason, in the main text we show examples with integration times only in the range of one period.

Finally, we emphasize that, according to Chua's own definition \cite{Chua1976}, a memristive device is indistinguishable from a nonlinear resistor at very low frequencies, and reduces to a linear resistor at high frequencies. This is exactly the case of our photonic quantum memristor, as shown in Eq. \eqref{eq_lowfreq} and \eqref{eq_hifreq}. 

For the practical implementation of Eq. \eqref{eq_integration2} we discuss how the feedback loop retrieves $\braket{n_\text{in}}$. This is straightforward, as the input is linked to the measurement in port D by  $\braket{n_\text{in}} = \braket{n_\text{meas,D}}/R$ (see Fig. \ref{fig_setup}a). In practice, we use a microcontroller that samples $\braket{n_\text{meas,D}}$, estimates $\braket{n_\text{in}}$ using the previous value of $R(t)$, performs the integral and consequently updates the value of $R(t')$. This operation is executed at a sampling rate of approximately 20 kSa/s. A control is implemented in the code that prevents $R(t)$ from going exactly to zero, otherwise the feedback would break.
\subsection{Experimental setup}
\label{sec_setup}

\begin{figure*}
    \centering
    \includegraphics[width=\textwidth]{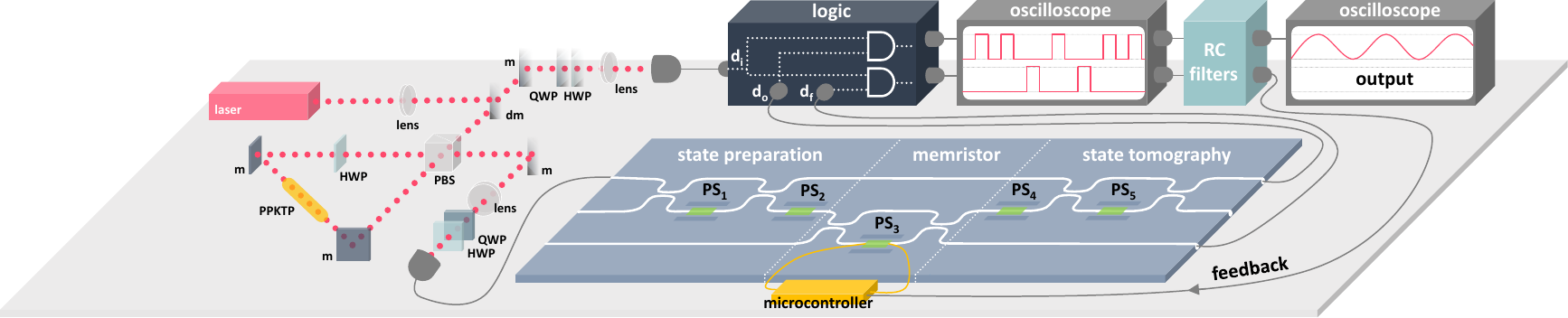}
    \caption{\textbf{Experimental setup.} Pumped by a continuous wave laser, a spontaneous parametric downconversion source using  periodically-poled KTP crystal (PPKTP) generates orthogonally polarised photon pairs with a wavelength of 1550 nm. By using a polarization beamsplitter (PBS) the photon pairs are separated such that the measurement of one photon heralds the presence of the other photon that is coupled into the photonic quantum memristor processor. The state preparation stage, consisting of a Mach-Zehnder interferometer with two tunable phase shifters (PS) allows for the preparation of the input state for the quantum memristor. In the final part of the device, the photon undergoes a tomography stage, which is used for characterisation purpose and otherwise set to perform an identity transformation. A logic unit analyses the coincidences between the idler photon and the output and feedback photons from the outputs of the photonic processor, triggering the emission of square pulses each time a coincidence is detected. By low-pass-filtering each channel with RC filters we obtain a measurement of photon number both for the output and the feedback signals.}
    \label{fig_fullsetup} 
\end{figure*}

The experimental setup used for our demonstration is detailed in Fig. \ref{fig_fullsetup}. A collinear Type II SPDC source emits pairs of identical photons at 1550 nm. The source is based on a 30 \si{\milli\metre} PPKTP crystal with a poling period of 46.2 \si{\micro\metre} adapted for downconversion from 775 to 1550 \si{\nano\metre}. The crystal is pumped by a CW amplified diode laser (Toptica TA Pro 780) with a pump power of approximately 80 \si{\milli\watt}. The crystal is inserted in a Sagnac interferometer which produces polarisation-entangled photons, although in this specific case the entanglement is not used. 

One of the photons (idler) is sent directly to the detectors for heralding, while the other (signal) is coupled to the integrated photonic processor via a single-mode fiber which is directly glued to its surface. In the photonic processor, the photon goes through the state preparation stage and then through the quantum memristor. At the output, the processor is pigtailed to single-mode fibers attached to the detectors. We use superconducting nanowire single-photon detectors (PhotonSpot Inc.) with average detection efficiency above 95\%. We use three detectors: one for the heralding (idler) photon, one for the feedback signal, and one for the output signal. 

After the detectors, a logic unit analyses the signals. Every idler-feedback coincidence triggers the generation of a square voltage pulse in the feedback channel. Equally, every idler-output coincidence triggers a voltage pulse in the output channel. With a pump power of about 80 mW in the source, the maximum coincidence rate in each channel is approximately $3 \times 10^4$ counts/s. Both channels are then low-pass filtered with $RC = 100$ ms filters. This has the effect of averaging the trains of pulses and providing a continuous voltage signal that is proportional to the pulse rate, which is in turn proportional to the photon number. Measuring the output voltage of the RC filters constitutes therefore a measurement of the photon number expectation value. Note that, following the discussion around Eq. \eqref{eq_integration2} in the main text, the time constant $RC$ should be much smaller than the integration time $T$, which is clearly the case as we used $T=10$ s.

At this point, the output signal goes to an oscilloscope for final data logging, while the feedback signal goes to a microcontroller which computes $\braket{n_\text{in}}$ and uses it to update the new value of $R(t)$ in the chip.

The system is then tested by varying the input number of photon as a sine wave. This is performed by a function generator which acts on the the integrated quantum state preparation stage.

\subsection{Quantum reservoir computing}

\label{sec_QRC}
%The topology of the machine learning network as a hybrid quantum/classical computing platform is mainly given in Fig. \ref{fig_newnet} though we give a generally pedagogical overview of the approach used for the results presented here. 
The input to the quantum reservoir is an amplitude-encoded quantum state on a $d$-dimensional complex Hilbert space. For $m$ modes and $p$ photons, the dimension $d$ grows combinatorially as ${m+p-1 \choose p}$ with this number giving the maximum dimension of classical input vectors. A classical input vector $\vec{v}\in \mathbb{R}^{n}$ is continuously encoded into a quantum state $q_i$ as $q_i\coloneqq \frac{1}{||v||_2}\sum_{j=1}^n  v_{j} \ket{j}$, where $\ket{j}$ is an eigenvector of the computational basis, with $n\leq d$ and $v_j$ the $j$th element of $\vec{v}$. This state is fed into a beamsplitter matrix with randomly set reflectivity for distributing the coherent quantum information across all modes and subsequently fed into $\floor{\frac{m}{3}}$ quantum memristors. The output of each memristor is used as input for another array of beamsplitters that enable a coherent interference after the nonlinear map due to the quantum memristors. We generally assume that the encoding and measurement rails for a quantum memristor do not overlap, thus for $M$ parallel optical memristors one requires $3M$ optical modes. For a full description one could compute the process tensor\! \cite{Pollock2018} over the time series length, however it is sufficient for us to consider a list of completely positive trace preserving maps that take the input quantum state and update the next map (dependent on the quantum memristor's state). The first map acts on the initial input state and the measurement outcomes of the quantum memristors adapt the next implementation map that acts on the next input to the reservoir. This is repeated $t$ times, where $t$ is the length of the time sequence for the input data.

The $t$th output density operator of the quantum reservoir is treated as the correlated output of the reservoir. This final output state is measured via a positive operator-valued measure (POVM) using the Fock basis $F$, with elements defined as $F_{ij\dots k}=\outerproduct{ij\dots k}{ij\dots k}$. A probability vector corresponding to the outcome probabilities of these projectors $\{ \tr[\rho_t F_{p00\dots0}],\tr[\rho_t F_{(p-1)10\dots0}], \tr[\rho_t F_{}] \dots, \tr[\rho_t F_{ij\dots k}]  \} \in \mathbb{R}^{d}$ serves as the training data for the classical readout network. 
%In practice this evolution would be done in a physical system and indeed it is the simulation of this process that the bulk of our computational efforts are spent. The training of the readout network itself, as a comparatively tiny one, is rapid, on the order of a minute.

Since the statistics of the photon counters are dependent on the quantum memristor's beam splitter settings of the previous iteration, the overall final output vector is correlated with the entire input sequence. In this way, correlations across time are propagated within the quantum reservoir and affect the output probability vector. Note, that the exact input state to the quantum memristors is no longer known, instead the memory effect is implemented as a discrete-time update rule analogous of Eq.\! \eqref{eq_mem2}, but with time steps defined as a set number of channel uses (generally around a thousand photon detection events each). This preserves the memory effect of the quantum memristors and leads to a stateful quantum reservoir that can be exploited for learning tasks both quantum and classical in nature.

The readout network itself is near trivial by design, as we want to ensure any sophisticated estimation is performed by the reservoir. The network exists purely as a linear mapping between the unknown outputs of the quantum reservoir and a human readable classifier. The output probability vector from the quantum reservoir is fed into the input layer and propagated through the network, which contains only a single hidden layer. All neurons except those in the output layer do not have an activation function and thus the neural network implements an entirely linear transform. 
%The final nonlinear layer is required because our task is commensurate with answering a binary (for entanglement detection) or trinary question (for image classification) and requires a definitive single answer. Thus the output function needs to be either clipped, normalised or scaled in some way that produces a probability vector for the networks estimate of the output class while still retaining the output relation with one another for gradient back-propagation. This amounts to a rescaling function that continuously maps the set of all reals $[-\infty,\infty ] \rightarrow [0, 1]$ as well as producing a categorical probability vector wherein the sum of the output is unity. Such behaviour is fulfilled by 
A softmax function operates on the $W$ neurons of the output layer of the neural network $\vec{q} \in \mathbb{R}^W$ as:
\begin{equation}
    s(\vec{x}) = \sum_{i}^W\frac{\exp(q_i)}{\sum_j \exp(q_j)} \vec{e}_i,
\end{equation}
and is only needed to map such output into a human-readable probability vector. The final output of the classifier is considered to be the largest of these probability values. No additional computational power comes from this nonlinear map as it is simply a function on the final output vector of the neural network. 
%Indeed, one could consider the true output of the network to be the vector $\vec{q}$ and the softmax function a transformation to make it human readable. 

\paragraph*{Image classification by sequential data analysis.} The first task discussed was a classical image classification problem. Both the training and testing sets consist of images of handwritten digits zero, three and eight, which are chosen for their similarity\! \cite{lecun2010mnist, tensorflow2015-whitepaper}. Each image is composed of $18\times 12$ pixels whose columns serve as the input to the quantum reservoir, encoded using the above amplitude encoding for each column and then sequentially fed into the network using a finite number of samples. Generally a thousand instances for each encoded state was found to be sufficient to gather approximately correct statistics, yielding the aforementioned 95\% accuracy. 

%In practice, the code \cite{Morris2021} performs the computationally intensive simulation of the quantum reservoir once, and then saves the result locally so that future runs (with perhaps tweaked learning parameters of the interpreter network) can be quickly tested. As long as the quantum optical parameters remain unchanged, the same simulation data may be freely reused for greatly reducing compute time.

%Though we have endeavoured to be thorough in commenting the code, readers are welcome to contact the authors for assistance in understanding or extending the work done already. In particular, the code used for simulating quantum optics and machine learning advances inspired by quantum optics, are standalone works and may be of interest aside from their usage here.

%Finally, although the results of the paper are initially encouraging, it is not obvious that the high performance seen is primarily due to any quantum effects. Certainly reservoir computers have been constructed that perform classification to a high degree of accuracy and so we have endeavoured to provide evidence that in the absence of quantum phenomena, the same network experiences either a serious performance drop or fails entirely.

For comparing the cases of quantum and classical reservoir computing we have chosen as strategy to keep the topology fixed (meaning optics and readout layer). This allows to switch between the quantum and classical cases by only varying a) how the data is encoded and b) how the reservoir's output is measured. The quantum case is explained above and we give a brief description of the classical simulation here. Instead of encoding data into quantum states, we use finitely bound coherent states that tend to the classical limit.

For input data assumed to be a real vector $\vec{v} \in \mathbb{R}^n$ of length $n$, we define a set of $n$ coherent states as 
\begin{equation}
\ket{\psi_k} = e^{-\frac{k^2}{2}} \sum_{n}^d \frac{k^n}{\sqrt{n!}}\ket{n},
\end{equation}
on an underlying $d$-dimensional Fock space (taking a finite truncation of the normally infinite series). This gives a list of $n$ encoding input states that we form a statistical mixture of as $c_i = C_i\sum_j v_j \outerproduct{\psi_j}{\psi_j}$, $C_i$ being a normalisation constant such that $\tr[c_i]$ is unity. This classical mixture $c_i \in \mathbb{B}(\mathcal{H}_d)$ serves as the 'classically' encoded input state, sequenced the same as it is for the quantum case. 
%This is a statistical mixture that only makes sense in the regime of repeated state input which remains the same as it does for the quantum case, though there we repeatedly generate single photons and manipulate them to the desired state, here we sample the $\ket{\psi}_i$ state with probability proportional to $v_i$. 
After propagation through the quantum reservoir, projective measurements are performed similar to the quantum case a discrete number of times to produce a sample probability that is then used to update the quantum memristor, ready for the next encoded state of the input sequence. We then see a considerable drop in the overall classification performance, thus giving strong evidence that the performance we are seeing is due to quantum effects. 

Furthermore, disabling the quantum memristor's update ability for the original scheme (with single photons) leads to a complete failure. This is unsurprising given that in this case the network must essentially classify the digit with only the very last column to work with - one that very often contains no significant information about the image. %Using different columns for the single input would likely yield improved performance however this detracts from the main point of having no memory - with it, classification is successful and so clearly some historical information is being propagated through the network and this information is key to the performance of the network. Thus, the quantum memristor with its classical memory is indeed playing an important role in the execution of the learning task.  

\paragraph*{Entanglement detection.} The second task considered was entanglement detection, where we require both, random entangled and separable pure states with respect to a bipartition on $\mathcal{H}_d \otimes \mathcal{H}_d$ for training, where $\mathcal{H}_d$ is the d-dimensional complex Hilbert space of the quantum reservoir. This is easily achievable as one can show that separable states are measure zero on the underlying set and so uniform random sampling over $\mathcal{H}_d \otimes \mathcal{H}_d$ will almost surely generate states belonging to the entangled class. For the non-entangled states we randomly sample two quantum states on $\mathcal{H}_d$ and compute their tensor product. Embedding these states into the larger input space of the quantum reservoir, the task is to identify if the input contains any entanglement with respect to the chosen bipartition, yielding the aforementioned 98\% success rate on never before seen states. 

The code for both tasks is available at \cite{Morris2021}.

\subsection{From the "original" memristor to a photonic memristor}
\label{sec_photmem}

\begin{figure}
\includegraphics[width=0.7\columnwidth]{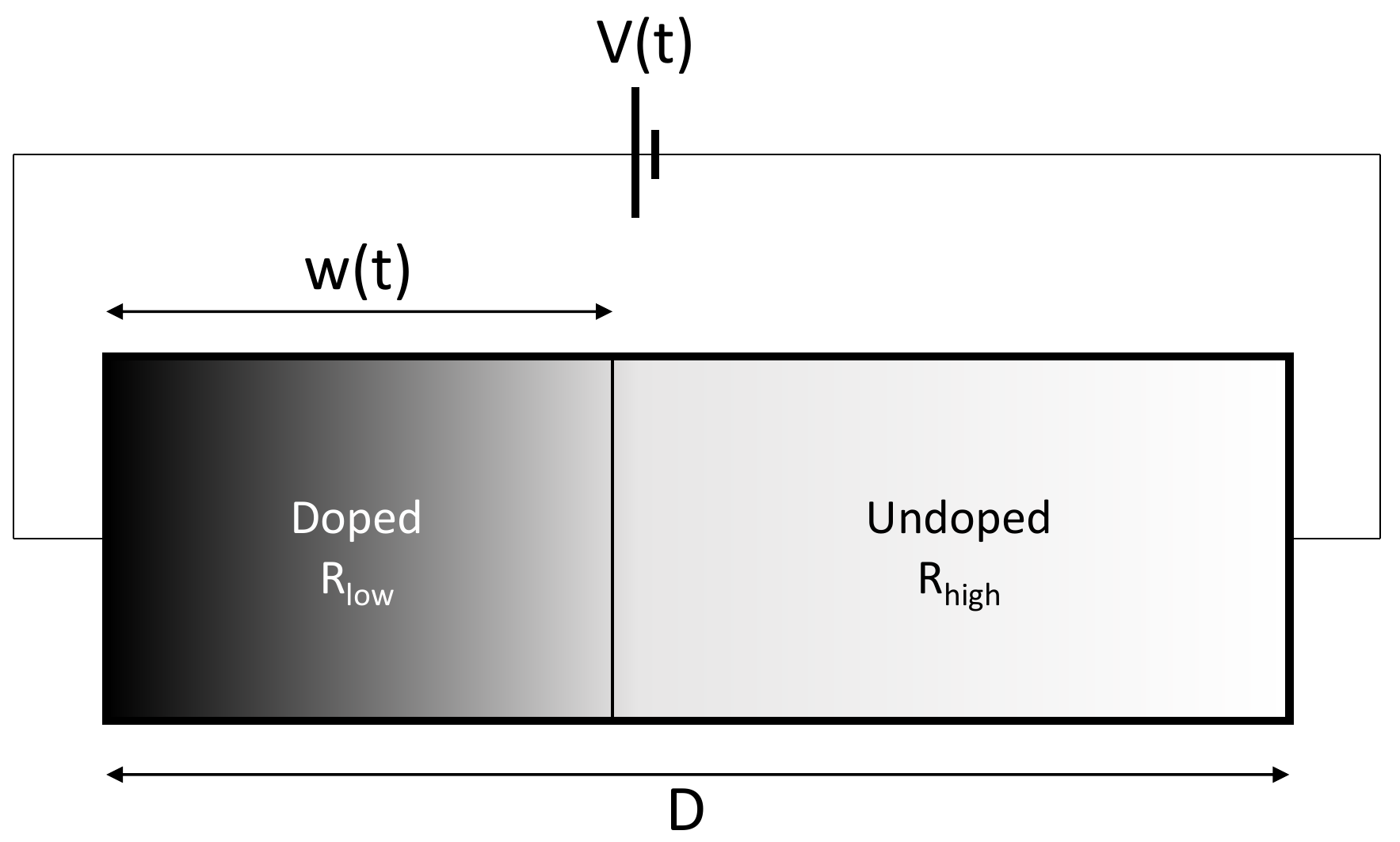}
\caption{\textbf{Original memristor by Struckov et al \cite{Strukov2008}.} When a voltage is applied across the junction, the ideal separation line between doped and undoped semiconductor shifts, thereby changing the resistance of the junction itself, and originating the hysteresis.}
\label{fig_originalmemristor}
\end{figure}

After being forgotten for several decades, the memristor had an explosive comeback in 2008 when Struckov et al. \cite{Strukov2008} reported for the first time a physical model of the "missing memristor" in a simple semiconductor junction. They considered a nanometric junction between a doped semiconductor with resistivity $R_\text{low}$ and thickness $w$, and an instrinsic semiconductor with resistivity $R_\text{high}$ and thickness $D-w$, as represented in Fig. \ref{fig_originalmemristor}. The overall resistance of the junction is easily calculated as the series of its two parts. However, when a voltage is applied to the junction it causes the ions of the doped part to drift, thus effectively changing the resistance of the junction itself. This can be modelled as a shift of $w$, so that the behaviour of this device can be described by the coupled equations
\begin{align}
	 	v(t) &= \bigg[R_\text{low}\frac{w(t)}{D}+R_\text{high}\bigg(1-\frac{w(t)}{D}\bigg)\bigg] \cdot i(t) \label{eq_ormem1}, \\
	 	\dot{w}(t) &= \mu\frac{R_\text{high}}{D}\cdot i(t), \label{eq_ormem2}
\end{align}
where $\mu$ is a constant related to the mobility of the ions inside the semiconductor. One can immediately see that these equations satisfy the form required by Eqs. \eqref{eq_mem1} and \eqref{eq_mem2}. Abstractly, one can think of this device as a sort of sliding potentiometer where the position of the slider $w(t)$ --- i.e. the state variable of the device --- is influenced by the past current that went through the device, hence the hysteresis.

\begin{figure}
    \includegraphics[width=\columnwidth]{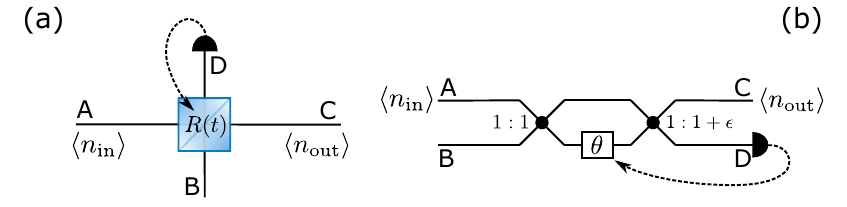}
    \caption{\textbf{Photonic quantum memristor.} a) Basic scheme of the photonic quantum memristor. b) The integrated optics equivalent, where the reflectivity is set by the phase $\theta$. The black lines correspond to guided modes of the integrated chip, and the crossing points are directional couplers, which act as beam splitters.}
    \label{fig_ortophotmem}
\end{figure}

Let us now consider the tunable beam splitter of Fig. \ref{fig_ortophotmem}a. In an idealized picture, one has the elementary relation
\beq 
\tag{\ref{eq_mem3}}
\braket{n_\text{out}} = (1-R(\theta))\braket{n_\text{in}},
\eeq 
where $R(\theta)=[0,1]$. In a real-world device, however, the reflectivity can never reach \emph{exactly} zero or one. That is because, even if one assumes a lossless device, a fraction of the input light (however small) will always leak into the undesired output arm. This is especially true in integrated optics, where a common way to realise a tunable beam splitter is by a Mach-Zehnder interferometer with a phase shifter in one of the arms, as shown in Fig. \ref{fig_ortophotmem}b. If the two directional couplers composing the Mach-Zehnder have a perfectly balanced splitting ratio of 1:1, the device has the ideal operation of Eq. \eqref{eq_mem3}. In practice, however, the exact splitting ratio can only be achieved up to a given experimental accuracy. Let us then consider a Mach-Zehnder where, for simplicity, one of the couplers is assumed to be perfectly balanced and the other to be slightly off. In the Heisenberg picture, the action of the Mach-Zehnder can be described by the matrix product of its three components:
\beq
\label{eq_mzmatrix1}
\begin{pmatrix}
\hat{a}_\text{C} \\ \hat{a}_\text{D}
\end{pmatrix}
=
\begin{pmatrix}
t & ir \\ ir & t
\end{pmatrix}
\begin{pmatrix}
1 & 0 \\ 0 & e^{i\theta}
\end{pmatrix}
\frac{1}{\sqrt{2}}
\begin{pmatrix}
1 & i \\ i & 1
\end{pmatrix}
\begin{pmatrix}
\hat{a}_\text{A} \\ \hat{a}_\text{B}
\end{pmatrix},
\eeq
where $\hat{a}$ is the annihilation operator on each spatial mode and the three matrices from left to right refer to the action of the unbalanced directional coupler, the phase shifter and the balanced coupler, respectively. Here, $r$ and $t$ indicate the reflection and transmission coefficients of the unbalanced coupler. By computing the product one obtains

\beq
\label{eq_mzmatrix2}
\begin{pmatrix}
\hat{a}_\text{C} \\ \hat{a}_\text{D}
\end{pmatrix}
= \frac{1}{\sqrt{2}}
\begin{pmatrix}
t-re^{i\theta} & i(t+re^{i\theta}) \\ i(t+re^{i\theta}) & -r+te^{i\theta}
\end{pmatrix}
\begin{pmatrix}
\hat{a}_\text{A} \\ \hat{a}_\text{B}
\end{pmatrix}.
\eeq

In analogy with the beam splitter of Fig. \ref{fig_ortophotmem}a, let us consider A as input port and C as output port, while input B is not used. One can extract from the previous matrix
\beq
\label{eq_mzmatrix3}
\hat{a}_\text{C} = \frac{1}{\sqrt{2}}(t-re^{i\theta}) \, \hat{a}_\text{A},
\eeq
from which we obtain
\beq
\label{eq_mzmatrix5}
\braket{n_\text{out}} = \frac{1}{2}(1-2rt\cos\theta) \braket{n_\text{in}}.
\eeq
In the ideal case, when $r=t=1/\sqrt{2}$, the equation reduces to the well known
\beq
\label{eq_idmz}
\braket{n_\text{out}} = \frac{1}{2}(1-\cos\theta) \braket{n_\text{in}},
\eeq
and by comparison with Eq. \eqref{eq_mem3} one easily finds
\beq
\label{eq_refl}
R(\theta) = \frac{1}{2}(1+\cos\theta).
\eeq
Because of the relation $r^2+t^2=1$, the product $rt$ in Eq. \eqref{eq_mzmatrix5} takes its maximum value of $1/2$ for $r=1/\sqrt{2}$. In a more realistic picture, where $r$ can only approximate the ideal value, one has $rt = 1/2-\eta$ with $\eta\ll 1$. Introducing this product in Eq. \eqref{eq_mzmatrix5} one obtains
\beq
\label{eq_mzmatrix6}
\braket{n_\text{out}} = \frac{1}{2}(1-(1-2\eta)\cos\theta) \braket{n_\text{in}}.
\eeq
By the definition of Eq. \eqref{eq_refl} and some simple algebra, one can rewrite the last equation as
\beq
\label{eq_bsreal}
\braket{n_\text{out}} = \big[\eta R(\theta) + (1-\eta)(1-R(\theta))] \braket{n_\text{in}},
\eeq
which is our final point. This equation is a reasonable representation of a real-world beam splitter, for which the transmission never reaches exactly zero or one. Indeed, one can only control $R(\theta)$, and for $R(\theta)=0$ the transmission is $1-\eta$, while for $R(\theta)=1$ the transmission is $\eta$. We call $\eta$ a \emph{leakage factor}, as it represents the fraction of photons that, regardless of the external control, always leaks into the undesired output.

At this point, we observe that Eq. \eqref{eq_bsreal} is formally identical to Eq. \eqref{eq_ormem1}, where $\braket{n_\text{in}}$ corresponds to $i(t)$, $\braket{n_\text{out}}$ corresponds to $v(t)$, $\eta$ corresponds to $R_\text{low}$, $1-\eta$ corresponds to $R_\text{high}$, and $R(\theta)$ corresponds to $w(t)/D$, both of them being adimensional quantities in the interval $[0,1]$ and acting as state variables for the respective device. We also emphasise that the analogy does not break down when considering an ideal beam splitter with $\eta=0$, as it just corresponds to an ideal memristor with $R_\text{low}=0$. This is nearly the case in reality, as the resistance of doped semiconductors is tipically many orders of magnitude smaller than intrinsic semiconductors, even at moderate doping. 

We have then found in the beam splitter a device that inherently replicates the form of Eq. \eqref{eq_ormem1}. Consequently, the natural choice of the feedback law would be one that replicates the dynamics of Eq. \eqref{eq_ormem2}. By the mere correspondence of the quantities, this would translate to $\dot{R}(\theta) \propto \braket{n_\text{in}}$. However, a feedback law of this type would be rather pointless because, unlike current and voltage, photon number can only take positive values and therefore $R(\theta)$ would just increase in time, eventually saturating to unity. One of the fundamental properties of a memristor, instead, is the ability to revert its state, which implies that $\dot{R}(\theta)$ must also take negative values. This can be readily obtained by a simple baseline shift:
\beq 
\tag{\ref{eq_mem4}}
\dot{R}(\theta,t) = \braket{n_\text{in}(t)} - 0.5\braket{n}_\text{max},
\eeq 
where $\braket{n}_\text{max}$ is the maximum value of $\braket{n_\text{in}}$ in time. In the single-photon case, $\braket{n_\text{max}}=1$, so the law effectively becomes $\dot{R}(\theta) = \braket{n_\text{in}} - 0.5$. It is easy to see that in such a case, input states with average number of photons lower than 0.5 will contribute negatively and bring $R$ to zero, while input states with average number of photons higher than 0.5 will contribute positively and bring $R$ to one. The resulting dynamics is an hysteresis figure that closely resembles that of the original memristor, though limited to the positive quadrant of the input-output plane.

\subsection{Quantum memristor for path-encoded photons}
\label{sec_comparison}

A quantum memristor in the photonic domain was first proposed by Sanz et al. \cite{Sanz2018}. Although setting an excellent groundwork, their scheme suffers from several drawbacks, most of which can be traced down to the choice of the x-quadrature operator as input variable of the device. 

In the example with Fock states --- which is arguably the most relevant for a quantum photonics application --- they obtain a hysteresis figure that is not pinched at the origin and thus, by their own definition, does not define a memristive device. This can be also seen from Eq. \eqref{eq_mem1}: when the input $u$ is zero the output $y$ has to be zero as well, so the hysteresis figure should invariably cross the origin. In their case this does not happen because an input state with zero quadrature does not always imply an output state with zero photons.

Most importantly, however, their scheme is very challenging to implement practically, as it requires the tuning and measurement of quadrature operators, which generally entails mixing the states with a coherent beam, thus greatly complicating any experimental setup. Furthermore, the input states are given by the superposition of Fock states, which is possible but impractical to realise in linear optics, especially when considering that the average quadrature of such a state depends on its relative phase term, which would have to be tightly controlled. An additional challenge is the fact that any subsequent manipulation of a qubit encoded in a superposition with the vacuum state would be highly nontrivial. 

Nevertheless, the paper by Sanz et al. does make an excellent point by highlighting that performing single-photon detection at one output of the beam splitter not only can be used in conjunction with classical feedback to produce a memory behaviour, but also has the effect of projecting the output on a coherent superposition of quantum states. Starting from this concept, we propose here a solution that substantially improves upon their scheme.

First, we use photon number as the input variable, rather than the quadrature. This is possible through a close formal analogy that we discovered between the equations of the beam splitter and the equations of the "original" memristor by Struckov et al. \cite{Strukov2008}, which we discussed in the previous section of the Supplementary Material. In addition to producing a hysteresis loop that always crosses the origin, using photon number as input variable allows us to drop the requirement of tuning and measuring quadrature operators, thus greatly simplifying the experimental footprint. 

Second, we address the challenge of creating and controlling superposition of Fock states by using a different encoding. In essence, we switch from number-encoding (also referred to as single-rail) to path-encoding, which is perhaps the most natural form of encoding in quantum photonics. The equivalence of the scheme is explained in details in Appendix \ref{sec_chipdesign}.

\subsection{Memristive behaviour arising from the phase shifters}
\label{sec_lpfmemristor}

\begin{figure}
    \centering
    \includegraphics[width=\columnwidth]{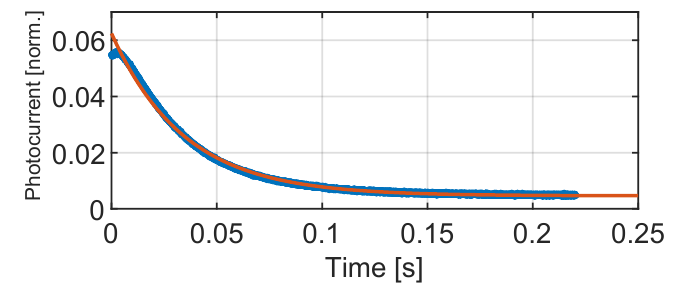}
    \caption{\textbf{Step response of the phase shifters.} Specifically, this shows the response of PS3 (see Fig. \ref{fig_setup}). We inject light in mode B and monitor the output power in mode C with a photodiode. When no voltage is applied to the phase shifter, the Mach-Zehnder is in cross state. When applying a step of $V_\text{bar} = 1.145$ V, the Mach-Zehnder switches to bar state, so the power in port C drops to zero. The resulting curve is well approximated by the step-response of a low-pass filter with cutoff frequency $f_\text{cut}=4.62$ Hz.}
    \label{fig_stepresp}
\end{figure}

The step response of the phase shifters in our chip is reported in Fig. \ref{fig_stepresp}. The curve is well approximated by a low-pass filter with a cutoff frequency of $f_\text{cut}=4.62$ Hz. If the memristor operates at frequencies $f_\text{osc} \ll f_\text{cut}$ the effect of the shifters is negligible. However, when approaching $f_\text{cut}$ the dynamics of the shifters starts to interfere with the dynamics of the feedback loop. To test the sole effect of the shifters, we set here our microcontroller to implement the identity $R(\theta) = \braket{n_\text{in}}$, rather than Eq. \eqref{eq_mem4}. At this point the dynamics of the device is only governed by how quickly the phase shifter can actually reach the value $R(\theta)$ set by the microcontroller. 

The resulting hysteresis figure, which we report in Fig. \ref{fig_memlpf}, is very similar to the one we obtained by implementing the windowed integrator (see Appendix \ref{sec_feedback}). This is not surprising, considering that a low-pass filter and a windowed integrator converge to the same limits both at low and high frequencies. Furthermore, this result indicates another viable path for the future development of these devices, where one could engineer the response of the thermal shifters to further simplify the quantum memristor layout by removing some components in the feedback signal processing.

\begin{figure*}
    \centering
    \includegraphics[width=\textwidth]{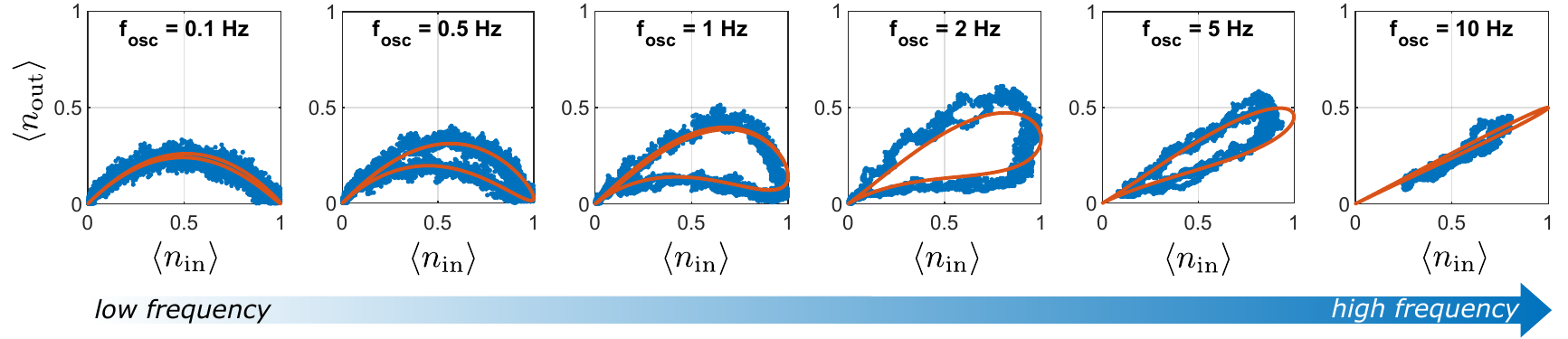}
    \caption{\textbf{Memristive behaviour originated by the phase shifters.} Simulations of a feedback loop containing a low-pass filter before the actuation of $R(\theta)$ (red lines) show good agreement with the experimental data (blue lines). The dynamics is very similar to the one we presented in Fig. \ref{fig_result}, which is unsurprising considering that low-pass filter and windowed integrator (see Appendix \ref{sec_feedback}) have a similar response. The data is more noisy because in this case we work with higher frequencies, so we have to use RC filters (see Appendix \ref{sec_setup}) with $RC = 10$ ms rather than 100 ms. Finally, the reason why at 10 Hz $\braket{n_\text{in}}$ does not cover the full [0,1] range is because the state preparation stage is controlled by an identical phase shifter, which also looses effectiveness when driven around or above its cutoff frequency.}
    \label{fig_memlpf}
\end{figure*}

\subsection{Experimental reconstruction of the output state's density matrix}
\label{sec_expdensmat}

The density matrix for the output state of our device was obtained in Appendix \ref{sec_chipdesign}:
\begin{equation}
    \tag{\ref{eq_densmat_dr}}
    \rho_\text{out,AB} =
    \begin{pmatrix}
    |\beta|^2R & 0 & 0 \\
    0 & |\alpha|^2 & \alpha^*\beta\sqrt{1-R} \\ 
    0 & \alpha\beta^*\sqrt{1-R} & |\beta|^2(1-R)
    \end{pmatrix}
\end{equation}
and is written in the basis $\ket{00}_\text{AB}, \ket{01}_\text{AB}, \ket{10}_\text{AB}$. Our chip features a tomography stage that allows in principle to fully reconstruct the output state at modes $A$ and $B$ (see Fig. \ref{fig_setup}e). The problem is that detecting photons at modes $A$ and $B$ automatically selects the $\ket{01}_\text{AB}$ or $\ket{10}_\text{AB}$ states, thus cancelling the $\ket{00}_\text{AB}$ term (i.e. the upper-left term of the density matrix).  In other words, one cannot use photon detection in $A$ and $B$ to characterise the absence of photons in $A$ and $B$. However, we know from Eq. \eqref{eq_rhoout_abc} that the term $|\beta|^2R$ corresponds to the fraction of photons going to mode $C$, that is the measurement port of the quantum memristor, connected to the feedback loop (see Fig. \ref{fig_setup}e). Therefore, by temporarily disconnecting the feedback loop, we can use the photon count at output $C$ to estimate $|\beta|^2R$. 

In essence, we first use the tomography stage at modes $A$ and $B$ to reconstruct the submatrix relative to the $\ket{01}_\text{AB}$ and $\ket{10}_\text{AB}$ terms, i.e. the lower-right 2x2 submatrix, using the most-likelihood method. We then use the number of detetected photons at output $C$ to estimate the upper-left term, and we obtain the final matrix by rescaling according to the normalisation $\Tr(\rho_\text{out,AB}) = 1$. 

On a final note, it is worth mentioning that the off-diagonal terms need to include an additional phase term $e^{i(\phi_\text{MZ} + \phi_\text{global})}$. Here, $\phi_\text{MZ}$ is the phase introduced by the Mach-Zehnder, which we know because we are controlling the reflectivity of the Mach-Zehnder, which is $R = \cos^2(\phi_\text{MZ}/2)$. On the other hand, $\phi_\text{global}$ is the global phase that originates in the chip by the difference in length of paths $A$ and $B$. This cannot be measured a priori, so we actually retrieve this term by fitting the phase of the off-diagonal terms to our data, obtaining $\phi_\text{global} = 5.6$ rad. 

Clearly, the output density matrix depends on the specific settings of $|\beta|^2$ and $R$ ($\alpha$ depends on $\beta$ because of the normalisation $|\alpha|^2 + |\beta^2| = 1$). We characterised the density matrix for the combinations of $|\beta|^2 = [0, 0.3, 0.7, 1]$ and $R = [0, 0.3, 0.5, 0.7, 1]$. Our results are summarised in Table \ref{fig_tomo_results}.

\begin{table*}
 \begin{tabular}{|c|c c || c c c | c c|} 
 \hline 
 & $|\beta|^2$ & $R$ & $\rho_\text{out,TH}$ & $\rho_\text{out,EXP}$ & Fidelity [\%] & $\Tr({\rho_\text{out,TH}}^2)$ & $\Tr({\rho_\text{out,EXP}}^2)$ \\
 \hline \hline
\textbf{1} & 0.0 & 0.0 & $ \begin{pmatrix}
        0.00 & 0 & 0 \\
        0 & 1.00 & -0.00-0.00i \\
        0 & -0.00+0.00i & 0.00 \\
        \end{pmatrix}$ & $\begin{pmatrix}
        0.00 & 0 & 0 \\
        0 & 1.00 & -0.02+0.03i \\
        0 & -0.02-0.03i & 0.00 \\
        \end{pmatrix}$ & 99.62 & 1.00 & 0.99 \\

        \textbf{2} & 0.3 & 0.0 & $ \begin{pmatrix}
        0.00 & 0 & 0 \\
        0 & 0.70 & -0.36-0.29i \\
        0 & -0.36+0.29i & 0.30 \\
        \end{pmatrix}$ & $\begin{pmatrix}
        0.00 & 0 & 0 \\
        0 & 0.70 & -0.33-0.27i \\
        0 & -0.33+0.27i & 0.30 \\
        \end{pmatrix}$ & 97.19 & 1.00 & 0.95 \\

        \textbf{3} & 0.3 & 0.3 & $ \begin{pmatrix}
        0.09 & 0 & 0 \\
        0 & 0.70 & -0.12-0.37i \\
        0 & -0.12+0.37i & 0.21 \\
        \end{pmatrix}$ & $\begin{pmatrix}
        0.09 & 0 & 0 \\
        0 & 0.70 & -0.09-0.36i \\
        0 & -0.09+0.36i & 0.21 \\
        \end{pmatrix}$ & 98.73 & 0.84 & 0.81 \\

        \textbf{4} & 0.3 & 0.5 & $ \begin{pmatrix}
        0.15 & 0 & 0 \\
        0 & 0.70 & -0.03-0.32i \\
        0 & -0.03+0.32i & 0.15 \\
        \end{pmatrix}$ & $\begin{pmatrix}
        0.15 & 0 & 0 \\
        0 & 0.70 & -0.01-0.31i \\
        0 & -0.01+0.31i & 0.15 \\
        \end{pmatrix}$ & 99.33 & 0.74 & 0.73 \\

        \textbf{5} & 0.3 & 0.7 & $ \begin{pmatrix}
        0.21 & 0 & 0 \\
        0 & 0.70 & 0.03-0.25i \\
        0 & 0.03+0.25i & 0.09 \\
        \end{pmatrix}$ & $\begin{pmatrix}
        0.22 & 0 & 0 \\
        0 & 0.70 & 0.05-0.24i \\
        0 & 0.05+0.24i & 0.08 \\
        \end{pmatrix}$ & 99.69 & 0.67 & 0.66 \\

        \textbf{6} & 0.3 & 1.0 & $ \begin{pmatrix}
        0.30 & 0 & 0 \\
        0 & 0.70 & 0.00-0.00i \\
        0 & 0.00+0.00i & 0.00 \\
        \end{pmatrix}$ & $\begin{pmatrix}
        0.31 & 0 & 0 \\
        0 & 0.69 & -0.02+0.03i \\
        0 & -0.02-0.03i & 0.00 \\
        \end{pmatrix}$ & 99.71 & 0.58 & 0.57 \\

        \textbf{7} & 0.7 & 0.0 & $ \begin{pmatrix}
        0.00 & 0 & 0 \\
        0 & 0.30 & -0.36-0.29i \\
        0 & -0.36+0.29i & 0.70 \\
        \end{pmatrix}$ & $\begin{pmatrix}
        0.00 & 0 & 0 \\
        0 & 0.31 & -0.29-0.29i \\
        0 & -0.29+0.29i & 0.69 \\
        \end{pmatrix}$ & 94.92 & 1.00 & 0.91 \\

        \textbf{8} & 0.7 & 0.3 & $ \begin{pmatrix}
        0.21 & 0 & 0 \\
        0 & 0.30 & -0.12-0.37i \\
        0 & -0.12+0.37i & 0.49 \\
        \end{pmatrix}$ & $\begin{pmatrix}
        0.21 & 0 & 0 \\
        0 & 0.31 & -0.06-0.36i \\
        0 & -0.06+0.36i & 0.48 \\
        \end{pmatrix}$ & 97.75 & 0.67 & 0.64 \\

        \textbf{9} & 0.7 & 0.5 & $ \begin{pmatrix}
        0.35 & 0 & 0 \\
        0 & 0.30 & -0.03-0.32i \\
        0 & -0.03+0.32i & 0.35 \\
        \end{pmatrix}$ & $\begin{pmatrix}
        0.35 & 0 & 0 \\
        0 & 0.31 & 0.02-0.32i \\
        0 & 0.02+0.32i & 0.34 \\
        \end{pmatrix}$ & 99.20 & 0.55 & 0.54 \\

        \textbf{10} & 0.7 & 0.7 & $ \begin{pmatrix}
        0.49 & 0 & 0 \\
        0 & 0.30 & 0.03-0.25i \\
        0 & 0.03+0.25i & 0.21 \\
        \end{pmatrix}$ & $\begin{pmatrix}
        0.49 & 0 & 0 \\
        0 & 0.30 & 0.06-0.24i \\
        0 & 0.06+0.24i & 0.21 \\
        \end{pmatrix}$ & 99.68 & 0.50 & 0.50 \\

        \textbf{11} & 0.7 & 1.0 & $ \begin{pmatrix}
        0.70 & 0 & 0 \\
        0 & 0.30 & 0.00-0.00i \\
        0 & 0.00+0.00i & 0.00 \\
        \end{pmatrix}$ & $\begin{pmatrix}
        0.71 & 0 & 0 \\
        0 & 0.29 & -0.01+0.03i \\
        0 & -0.01-0.03i & 0.00 \\
        \end{pmatrix}$ & 99.63 & 0.58 & 0.59 \\

        \textbf{12} & 1.0 & 0.0 & $ \begin{pmatrix}
        0.00 & 0 & 0 \\
        0 & 0.00 & -0.00-0.00i \\
        0 & -0.00+0.00i & 1.00 \\
        \end{pmatrix}$ & $\begin{pmatrix}
        0.00 & 0 & 0 \\
        0 & 0.01 & 0.04-0.09i \\
        0 & 0.04+0.09i & 0.98 \\
        \end{pmatrix}$ & 98.45 & 1.00 & 0.99 \\

        \textbf{13} & 1.0 & 0.3 & $ \begin{pmatrix}
        0.30 & 0 & 0 \\
        0 & 0.00 & -0.00-0.00i \\
        0 & -0.00+0.00i & 0.70 \\
        \end{pmatrix}$ & $\begin{pmatrix}
        0.29 & 0 & 0 \\
        0 & 0.01 & 0.06-0.06i \\
        0 & 0.06+0.06i & 0.70 \\
        \end{pmatrix}$ & 98.87 & 0.58 & 0.59 \\

        \textbf{14} & 1.0 & 0.5 & $ \begin{pmatrix}
        0.50 & 0 & 0 \\
        0 & 0.00 & -0.00-0.00i \\
        0 & -0.00+0.00i & 0.50 \\
        \end{pmatrix}$ & $\begin{pmatrix}
        0.49 & 0 & 0 \\
        0 & 0.01 & 0.05-0.04i \\
        0 & 0.05+0.04i & 0.50 \\
        \end{pmatrix}$ & 99.07 & 0.50 & 0.50 \\

        \textbf{15} & 1.0 & 0.7 & $ \begin{pmatrix}
        0.70 & 0 & 0 \\
        0 & 0.00 & 0.00-0.00i \\
        0 & 0.00+0.00i & 0.30 \\
        \end{pmatrix}$ & $\begin{pmatrix}
        0.70 & 0 & 0 \\
        0 & 0.01 & 0.04-0.02i \\
        0 & 0.04+0.02i & 0.29 \\
        \end{pmatrix}$ & 99.24 & 0.58 & 0.58 \\

        \textbf{16} & 1.0 & 1.0 & $ \begin{pmatrix}
        1.00 & 0 & 0 \\
        0 & 0.00 & 0.00-0.00i \\
        0 & 0.00+0.00i & 0.00 \\
        \end{pmatrix}$ & $\begin{pmatrix}
        0.99 & 0 & 0 \\
        0 & 0.01 & 0.00-0.00i \\
        0 & 0.00+0.00i & 0.00 \\
        \end{pmatrix}$ & 99.05 & 1.00 & 0.98 \\
 \hline
\end{tabular}
\caption{\textbf{Tomography of the output states.} The results are obtained for several combinations of input states (determined by $|\beta|^2$) and reflectivity $R$ of the Mach-Zehnder. All the measurement are in excellent agreement with the theoretical values, showing an average fidelity $F=98.7$\%, and showing that no additional decoherence is significantly introduced by our device.}
\label{fig_tomo_results}
\end{table*}

%\end{appendices}
%TC:endignore
%%%%%%%%%%%%%%%%%%%%%%%%%%%%%%%%%%
%%%%%%%%%%%%%%%%%%%%%%%%%%%%%%%%%%

\end{document}